\documentclass{emulateapj}

\usepackage{amsmath}
\usepackage{graphicx}
\usepackage{multirow}
\usepackage{color}
\usepackage{natbib}

 at 10truept

\def\ffrac#1#2{{\textstyle\frac{#1}{#2}}}
\newcommand{\half}{{\textstyle{1\over2}}}

\shorttitle{Planets near mean-motion resonances}
\shortauthors{Petrovich, Malhotra \& Tremaine}
\begin{document}

\def\etal{et al.\ \rm}
\def\ba{\begin{eqnarray}}
\def\ea{\end{eqnarray}}
\def\etal{et al.\ \rm}
\def\sgn{\mbox{sgn}}
\def\rmk{\mbox{k}}

\newcommand{\mth}{m_{\rm th}}
\newcommand{\Mth}{M_{\rm th}}
\newcommand{\rs}{r_{\rm s}}
\newcommand{\phip}{\Phi_p}
\newcommand{\cs}{c_{\rm s}}
\newcommand{\mplanet}{m_{\rm pl}}
\newcommand{\ls}{l_{\rm sh}}
\newcommand{\third}{\rm third}
\newcommand{\yr}{\,\mbox{yr}}

\title{Planets near mean-motion resonances}

\author{Cristobal Petrovich\altaffilmark{1}, Renu
  Malhotra\altaffilmark{2}, \& Scott Tremaine\altaffilmark{3} }

\altaffiltext{1}{Department of Astrophysical Sciences, Princeton University, Ivy Lane, Princeton, NJ 08544, USA; 
cpetrovi@princeton.edu}
\altaffiltext{2}{Lunar and Planetary Laboratory, The University of Arizona, Tucson, AZ 85721, USA; renu@lpl.arizona.edu}
\altaffiltext{3}{School of Natural Sciences, Institute for Advanced
  Study, Einstein Drive, Princeton, NJ 08540, USA; tremaine@ias.edu}

\begin{abstract}
  The multiple-planet systems discovered by the {\it Kepler} mission
  exhibit the following feature: planet pairs near first-order
  mean-motion resonances prefer orbits just outside the nominal
  resonance, while avoiding those just inside the resonance.  We
  explore an extremely simple dynamical model for planet formation, in
  which planets grow in mass at a prescribed rate without orbital
  migration or dissipation. We develop an analytic version of this
  model for two-planet systems in two limiting cases: the planet mass
  grows quickly or slowly relative to the characteristic resonant
  libration time.  In both cases the distribution of systems in period
  ratio develops a characteristic asymmetric peak-trough structure
  around the resonance, qualitatively similar to that observed in the
  {\it Kepler} sample.  We verify this result with numerical
  integrations of the restricted three-body problem.  We show that for
  the 3\,:\,2 resonance, where the observed peak-trough structure is
  strongest, our simple model is consistent with the observations for
  a range of mean planet masses 20--$100M_\oplus$. This mass range is
  higher than expected, by at least a factor of three, from the few
  {\it Kepler} planets with measured masses, but part of this
  discrepancy could be due to oversimplifications in the dynamical
  model or uncertainties in the planetary
  mass-radius relation. 
\end{abstract}



\section{Introduction}  
\label{sect:intro}

\noindent
As of fall 2012, almost 100 multi-planet systems have been detected
by ground-based radial-velocity observations, and 365 transiting multi-planet
systems have been detected by the {\it Kepler}
spacecraft\footnote{Technically most of these are planet candidates,
  since they have not been confirmed by radial-velocity measurements,
  but the expected false-positive rate in multi-planet transiting
  systems is quite low \citep{liss12}.}. These large samples enable statistical studies of
correlations between the properties of members of multi-planet
systems. One result from these studies is that many planets appear to
be close to mean-motion resonances, that is, their orbital
periods are close to the ratio of two small integers (see below for a
more precise definition). In particular:

\begin{itemize}

\item \cite{LF11} comment that about one-third of the multi-planet systems
  studied by radial-velocity measurements contain near-resonant planet
  pairs, with about half of these near the 2\,:\,1 resonance. This is
  probably an underestimate because of a detection bias: there is an approximate degeneracy
  between the signal from an interior planet near the 2\,:\,1 resonance
  and harmonics arising from non-zero eccentricity in the outer planet
  orbit, which makes it difficult to detect low-mass planets in resonance with exterior
  companions \citep[e.g.,][]{ae10}.

\item \cite{FL12} find that {\it Kepler} planet pairs 
  with orbital period ratios within a few percent of 2\,:\,1 or 3\,:\,2 are
  preferentially found just wide of the resonance (i.e., period ratio slightly
  larger than 2 or 1.5) and tend to avoid spacings just narrow of the
  resonance.

\item Near-resonances involving three or more planets are also present in
  the {\it Kepler} data. \cite{LF11} find that the four planets in the
  system KOI-730 are in a chain of resonances, with period ratios
  8\,:\,6\,:\,4\,:\,3 to within 0.1\%. Lissauer et al.\ also find strong evidence
  for other resonances, including two three-planet resonances in the five-planet system
  KOI-500. \cite{FL12} find that three planets in the
  systems KOI-720 and KOI-2086 have mean motions $n$ that satisfy
  $5n_2-3n_4-2n_1\simeq0$ and
  $n_1-2n_2+n_3\simeq0$ respectively, to within 0.01\% (the
  planets are labeled in order of increasing period). 

\item In the GJ 876 system, planets 2, 3, and 4 have periods close to
  the ratio 1\,:\,2\,:\,4; planets 2 and 3 are close to a secular resonance;
  and there are other probable near-resonances (see \citealt{bal11}
  and references therein). Because of these near-resonances,
  gravitational interactions between the planets are detectable in the
  radial-velocity data, and these can be used to constrain the mutual
  inclinations and determine whether various critical arguments
  librate or circulate.

\item Within the solar system, Jupiter and Saturn are within 1\% of a
  5\,:\,2 resonance (the ``great inequality''), Uranus and Neptune are
  within 2\% of a 2\,:\,1 resonance, and Pluto is in a 3\,:\,2 resonance with
  Neptune.  There are also many resonances among the satellites of
  Jupiter and Saturn. Another notable near-resonance occurs between the two
  outer planets of the pulsar PSR B1257+12, which are within 2\% of a
  3\,:\,2 resonance. This near-resonance produces gravitational
  interactions large
  enough to be easily detectable, which allowed the existence of the
  planets to be confirmed and their inclinations to be measured
  shortly after their discovery \citep{ras92,mal92}.

\end{itemize}

Loosely speaking, an orbital resonance between two planets occurs when
their mean motions or orbital frequencies $n_1,n_2$ are nearly
commensurate, i.e., $n_1/n_2$ is close to a ratio of small
integers, $p:p+q$ where $p\not=0$ and $q\ge0$. The case $q=0$ is
sometimes called a corotation or co-orbital resonance; examples in the
solar system include the Trojan asteroids (1\,:\,1 resonance with Jupiter)
and the Saturnian satellites Janus and Epimetheus, but no extrasolar
co-orbital resonances are known.  When $q>0$ it is called the order of
the resonance, since for planets on nearly circular, coplanar orbits
the strength of the resonance potential is proportional to $e^q$ or
$I^q$ where $e$ and $I$ are the eccentricities and inclinations of the
resonant planets. Inclination resonances occur only for even $q$.

In a resonant configuration, the longitude of the planets at every
$q$th conjunction librates slowly about a direction determined by the
lines of apsides and nodes of the planetary orbits.  In the
action-angle variables for the Keplerian potential, this geometry is
naturally described by the libration of a so-called critical argument
which is a linear combination of the angle variables \citep{mal94,mal98,md99}.
 
\paragraph{Definition of resonance} Generally, a planet pair is
said to be ``in resonance'' if some dynamically significant critical
argument librates. If the critical argument circulates the planet pair
is said to be near but not in resonance. This definition has several
shortcomings for our purposes \citep[see for example][]{hl83,las12}:
(i) it is not always consistent with the expectation that resonant
planets have orbital periods close to the ratio of two small integers,
since planets can be resonant at {\em any} period ratio\footnote{As
  the eccentricity approaches zero, the apsidal precession rate due to
  a perturber grows without limit. Thus a critical angle
  whose time derivative involves the mean motions and precession rate
  can librate even when the mean motions are far from commensurability.};
(ii) by suitable canonical transformations one can change the
appropriate critical argument from libration to circulation, so this
definition of resonance is coordinate-dependent; (iii) interaction of
nearby resonances arising from the degenerate frequencies in the
Kepler problem can cause a critical argument to jump between libration
and circulation at irregular intervals.

Given these comments, it is nugatory to discuss whether {\it Kepler}'s
multi-planet systems are ``resonant'' or ``non-resonant'', and in this
paper we shall avoid the term ``resonant'' in favor of the looser
description ``near-resonant''.

\paragraph{Resonances and migration} Convergent
migration---evolution of the semi-major axis of one or both planets
such that the period ratio approaches unity---can lead to permanent
capture into resonance, and is believed to be the cause of the
Neptune-Pluto resonance \citep{mal93} and the resonances between
satellites of Jupiter and Saturn (although in these cases the
migration is outward, whereas migration in most exoplanet systems is
believed to be inward).

The existence of near-resonant planet pairs is often ascribed to
convergent migration \citep[e.g.,][]{spn01}, but there are problems with this
hypothesis: 

\begin{itemize}

\item The fraction of planets in resonance is quite small: \cite{FL12}
  find peaks at the 3\,:\,2 and 2\,:\,1 resonances of $\sim 20$ planet
  pairs each, out of a total sample of $\sim 750$ planet pairs; this
  small fraction is noteworthy since capture into the 2\,:\,1
  resonance is certain during convergent migration if the planets
  cross the resonance slowly enough and their initial eccentricities
  and inclinations are small enough. ``Slowly enough'' means a migration
  timescale larger than 
  $10^4\mbox{\,yr}(P_{\rm pl}/100\mbox{\,d})(10M_\oplus/m_{\rm pl})^{4/3}$ where
  $m_{\rm pl}$ and $P_{\rm pl}$ are the migrating planet's mass and orbital
  period; ``small enough'' means initial eccentricity smaller than
  $0.05(m_{\rm pl}/10M_\oplus)^{1/3}$ (see Appendix). Typical Type I
  migration times for {\it Kepler} planets in a low-mass
  protoplanetary disk are $10^3$--$10^4\mbox{\,yr}$ \citep{ttw02}, in
  which case migrating planets might avoid resonance capture according
  to this criterion, but orbit integrations by \cite{rein12b} that
  include eccentricity damping show that most {\it Kepler} planets are
  captured even if the migration time is as short as
  $10^3\mbox{\,yr}$.

\item Convergent migration typically
leads to capture into a 2\,:\,1 resonance. Capture into resonances with
smaller separations, such as 3\,:\,2 or 4\,:\,3, becomes increasingly
difficult: for example, to capture a planet into the 3\,:\,2 resonance
requires either that the initial conditions are fine-tuned so that the
planets form in the narrow interval between the 2\,:\,1 and 3\,:\,2 resonance
(only 13\% of the outer planet's semi-major axis), or that the
migration rate is fast enough that the planet jumps the weaker 2\,:\,1
and is captured at the 3\,:\,2 resonance. Yet \cite{FL12} find that the
excess of planet pairs near the 3\,:\,2 resonance is at least as strong as
the excess at the 2\,:\,1 resonance (see \citealt{rein12} for a
detailed numerical study of these issues). 

\item In the sample of {\it  Kepler} multi-planet systems examined
by \cite{FL12}, the excess of planet pairs at period ratios just
larger than 2 is accompanied by a deficit at period
ratios just smaller than 2. Both the peak and the trough have an
equivalent width of about 20 planets. This strongly suggests that the
features at this resonance arise from rearranging the periods of
planets near the resonance, rather than by capturing planets at the
resonance, which should produce a peak but no trough. 

\end{itemize} 

\cite{bm12} and \cite{lw12} have pointed out that dissipation due to
tides from the host star or the protoplanetary disk tends to repel
near-resonant planets, in the sense that their period ratios evolve
away from unity (see also \citealt{las12}). Thus, if migration is
common, dissipation could explain why there are so few near-resonant
planet pairs; and if there is little or no migration, dissipation
could explain why planet pairs that initially happen to lie near a
resonance are now preferentially found wide of the resonance. This
hypothesis is discussed further in \S\ref{sec:previous}. 

Although these and other analyses in the literature shed considerable
light on the behavior of near-resonant planets in the presence of
dissipation and/or migration, our view is that they put the cart
before the horse: the question that should be addressed first is, what
is the distribution of mean motions or period ratios expected near a
resonance in the {\em absence} of dissipation or migration? 

Efforts to address this question have a long and rich history, mostly
in the context of the asteroid belt, in which the distribution of mean
motions contains gaps at the 4\,:\,1, 3\,:\,1, 5\,:\,2, 7\,:\,3, and
2\,:\,1 resonances with Jupiter (the Kirkwood gaps) and peaks at the
3\,:\,2 and 1\,:\,1 resonances (the Hilda family and the Trojan
asteroids). Most of these features can be largely explained through
chaotic evolution of the asteroid orbits on timescales as long as 1
Gyr, long after the formation of the solar system was complete
\citep{wis83,Murray:1997,Lecar:2001}.
Although the effects of dissipation and migration are discernible
in the orbital distribution of asteroids, these features are comparatively
subtle \citep{Liou:1997,Minton:2009,Minton:2010}.
Thus the asteroid belt demonstrates that dissipation and large-scale
migration are not essential to produce near-resonant features in the mean-motion
distribution. Unfortunately for our purposes, the chaotic evolution of the
asteroids appears to depend strongly on the details of the planetary
configuration in the solar system and thus the insights gained from
studies of the asteroid belt cannot be immediately applied to exoplanets.

In this paper we focus on a simple and highly idealized model for the
formation of near-resonant features: at an initial time $t=0$ we place a
single planet and a large number of test particles on circular orbits
around the host star. The test particles are interior to the planet
and smoothly distributed in semi-major axis (see Equation
\ref{eq:epsilon}). The mass of the planet is $m_{\rm pl}h(t)$ where $h(t)$
ramps up from zero at time zero to unity at large times; we focus on
the two limiting cases in which the planet mass grows rapidly---$h(t)$
is a step function at $t=0$---and slowly compared to the
characteristic orbital and secular frequencies of the test particles;
see Equation (\ref{eq:tanh}).

\section{Analytic results}
\label{sect:analytic}

\subsection{Orbital elements}

\noindent
We follow the motion of the test particles using osculating Keplerian
orbital elements, with $a,e,\omega,\ell$ being the semi-major axis,
eccentricity, longitude of periapsis and mean anomaly.  For canonical
elements we use the modified Delaunay variables $\lambda=\ell+\omega,
\gamma=-\omega, \Lambda = (\mu a)^{1/2}, \Gamma = (\mu
  a)^{1/2}[1-(1-e^2)^{1/2}]$, where $\mu=Gm_*$ is the gravitational mass of
the star.

The Hamiltonian for the restricted three-body problem (star, planet,
test particle) is
\begin{align}
H(\lambda,\gamma,\Lambda,\Gamma,t) &=  -{\mu^2\over 2\Lambda^2} 
-\mu_{\rm pl}\left[{1\over{ |{\bf r}-{\bf r}_{\rm pl}|} }
-{ {\bf r}\cdot{\bf r}_{\rm pl} \over r_{\rm pl}^3}\right] \nonumber \\
&=  -{\mu^2\over 2\Lambda^2} + \mu_{\rm pl} H_{\rm pl}(\lambda,\gamma,\Lambda,\Gamma,t)
\label{e:eqn1}
\end{align}
where $\mu_{\rm pl}=Gm_{\rm pl}$ is the gravitational mass of the planet and ${\bf
  r},{\bf r}_{\rm pl}$ are the position vectors of the test particle and
planet relative to the star.  The first term describes the unperturbed
Keplerian motion of the test particle about the Sun, and the remainder
describes the perturbation from the planet.

Transit observations of exoplanets measure the time interval $P_{\rm
  tr}$ between successive transits, that is, the interval in which the
longitude increases by $2\pi$. We must relate the transit-based mean
motion $n_{\rm tr}\equiv 2\pi/P_{\rm tr}$ to the osculating mean
motion $n$, given by $n^2=\mu/a^3$. Write the perturbing Hamiltonian
as
\begin{equation}
H_{\rm pl}(\lambda,\gamma,\Lambda,\Gamma,t)=H_{\rm lp}(\gamma,\Lambda,\Gamma,t)
+ H_{\rm sp}(\lambda,\gamma,\Lambda,\Gamma,t) 
\label{e:eqnsp}
\end{equation}
where
\begin{equation}
H_{\rm lp}=\langle H_{\rm pl}\rangle_t, \quad H_{\rm   sp}=H_{\rm pl}-H_{\rm  lp},
\end{equation}
with $\langle\cdot\rangle_t$ denoting a time average over an interval
comparable to the total observational timespan; $H_{\rm sp}$ and
$H_{\rm lp}$ are respectively the short-period and long-period
 parts of the Hamiltonian. Then $n_{\rm tr}$ is the time average of the rate of change of the mean
longitude, $n_{\rm tr}=\langle d\lambda/dt \rangle_t$, and we have
\begin{equation}
n_{\rm tr}=\left\langle \frac{d\lambda}{dt}\right\rangle_t=\left\langle \frac{\partial
    H}{\partial\Lambda}\right\rangle_t=\frac{\mu^2}{\Lambda^3}+\frac{\partial
  H_{\rm lp}}{\partial\Lambda}=n+\frac{\partial  H_{\rm lp}}{\partial\Lambda}
\end{equation}
where $n$ is the osculating mean motion.

For the resonances we shall examine, the second term in these
equations is small; thus we may neglect the distinction between $n$
and $n_{\rm tr}$. In other words, the distribution of transit-based period
ratios near resonances is determined mainly by the distribution of the
osculating mean motions rather than by any differences between the
osculating and transit-based mean motions.

\subsection{Resonant dynamics}

\noindent
We consider the vicinity of a first-order mean-motion resonance, where
the ratio of the mean motions of the planet and the test particle is 
$p:(p+1)$. Thus $p>0$ corresponds to an interior resonance (test
particle inside the planet) and $p<-1$ is an exterior resonance. 

We now assume that the resonances are sufficiently well separated
compared to the size of the perturbations induced by the planet that
we can ignore all of the perturbations except those associated with
the $p:(p+1)$ resonance (we call this the single-resonance
approximation). It is
useful to make a canonical transformation to slow and fast variables,
\begin{eqnarray}
\phi &=& (p+1)\lambda_{\rm pl}-p\lambda +\gamma, \qquad\! \Phi = \Gamma; \cr
\psi &=& \lambda-\lambda_{\rm pl}, \quad\qquad\qquad\qquad \Psi = \Lambda+p\Gamma,
\label{e:eqn2}
\end{eqnarray}
where $\lambda_{\rm pl}=n_{\rm pl}(t-t_0)$ is the mean longitude of the planet, and
$n_{\rm pl}$ is its mean motion.  Then the new Hamiltonian is
\begin{align}
{\widetilde H} = &n_{\rm pl}[ (p+1)\Phi -\Psi] -{\mu^2\over2(\Psi-p\Phi)^2}
\nonumber \\
& \qquad +{\mu_{\rm pl}}H_{\rm pl}(\phi,\psi,\Phi,\Psi; a_{\rm pl},e_{\rm pl})
\label{e:htilde}
\end{align}
where $a_{\rm pl}$ is the semi-major axis of the planet's orbit and $H_{\rm pl}$
represents the planetary perturbation.  For simplicity we have assumed
that the planet is on a circular orbit and that the test-particle and
planet orbits are coplanar. Since $\psi$ is a fast variable, we will
drop $\psi$-dependent terms, which is equivalent to replacing $H_{\rm pl}$ by
$H_{\rm lp}$; consequently, the resonance Hamiltonian is independent of
$\psi$ and $\Psi$ is a constant of the motion.  We shall use the
notation
\begin{equation}
n_c = {\mu^2\over \Psi^3},\qquad a_c = {\Psi^2\over \mu};
\label{eq:stardef}
\end{equation}
$n_c$ and $a_c$ are constants of the motion which equal the osculating
mean motion and semi-major axis of the test particle when its
eccentricity is zero.

If the test-particle orbit is nearly circular, $\Phi\simeq
{1\over2}\sqrt{\mu a}e^2$ is small, and we can approximate
Equation (\ref{e:htilde}) with a few terms in an expansion in powers of
$\sqrt{\Phi}$,
\begin{equation}
{\widetilde H}_{\rm res} = [(p+1)n_{\rm pl}-pn_c]\Phi  +\beta \Phi^2 +\varepsilon \sqrt{2\Phi} \cos\phi,
\label{e:hres}
\end{equation}
where we have dropped an inessential constant, and 
\begin{equation}
\beta = -{3p^2n_c\over2\Psi}, \qquad
\varepsilon = -{\mu_{\rm pl}\over a_{\rm pl}}{f_p\over\sqrt{\Psi}}.
\end{equation}
Note that $\Psi\simeq\sqrt{\mu a}(1+\frac{1}{2}pe^2)$, and since the eccentricity is small, we have
$\Psi>0$ and $\beta<0$ in all cases of interest. The
coefficient $f_p$ is given by \citep{md99}
\begin{equation}
f_p=-(p+1+\ffrac{1}{2}D)b_{1/2}^{(p+1)}(\alpha),\quad
    \alpha=[p/(p+1)]^{2/3}
\end{equation}
when $p>0$ (planet exterior to test particle), and 
\begin{equation}
f_p=-\alpha(p+\ffrac{1}{2}-\ffrac{1}{2}D)b_{1/2}^{(|p+1|)}(\alpha)
-{\delta_{p,-2}\over2\alpha},\     \alpha=[(p+1)/p]^{2/3} 
\end{equation}
when $p<-1$ (planet interior to test particle).  Here $\delta_{i,j}$
is the Kronecker delta function, $D\equiv d/d\log\alpha$, and the Laplace coefficient
\begin{equation}
b_{1/2}^{(m)}(\alpha)=\frac{1}{\pi}\int_0^{2\pi}\frac{\cos
  mx\,dx}{(1-2\alpha\cos x +\alpha^2)^{1/2}}.
\end{equation}
We have
\begin{align}
f_1=-1.1905, \quad 
f_2=&-2.0252, \quad 
f_3=-2.8404;  \nonumber \\
f_{-2}=0.26987, \quad 
f_{-3}&=1.8957, \quad 
f_{-4}=2.7103. 
\end{align}

Following \citet{hl83}, we define a dimensionless time and canonical momentum, 
\begin{eqnarray}
\tau &=& \Bigg|{\beta\varepsilon^2\over4}\Bigg|^{1/3} t,\cr
R &=& \Bigg|{2\beta\over\varepsilon}\Bigg|^{2/3}\Phi,
\label{e:scale}
\end{eqnarray}
and a modified canonical coordinate $r$, 
\begin{eqnarray}
r &=& -\phi  \qquad\qquad \hbox{if $\varepsilon >0$} \cr
    &=& \pi -\phi \qquad\quad \hbox{if $\varepsilon <0$}.
\end{eqnarray}
The new, dimensionless Hamiltonian in the canonical variables $(r,R)$
is then given by 
\begin{equation}
K = -3\Delta R +R^2  - 2 \sqrt{2R} \cos r
\label{e:Kres}
\end{equation}
where the dimensionless resonance distance $\Delta$ is 
\begin{equation}
\Delta = {(p+1)n_{\rm pl}-pn_c \over |27\beta\varepsilon^2/4|^{1/3}}.
\label{eq:delta}
\end{equation}

A few notes:
\begin{itemize}

\item{} The strength of the resonance can be parametrized by the
  dimensionless ratio
  \begin{align}
s&\equiv \frac{|27\beta\varepsilon^2/4|^{1/3}}{pn_c}\nonumber \\
&=\sgn(p)\left|\frac{9f_pm_{\rm pl}a_c}{\sqrt{8|p|}m_* a_{\rm pl}}\right|^{2/3}.
\label{eq:sdef}
\end{align}
Note that $s$ is positive for interior resonances and negative for
exterior ones. 

\item{}  For small eccentricity, we can write $2R\simeq (e/s_e)^2$,
  where the eccentricity scale is 
\begin{align}
s_e \equiv& {1\over (\mu a_c)^{1/4}}\left|{\varepsilon\over
    2\beta}\right|^{1/3}\nonumber \\
=&\left|\frac{f_p m_{\rm pl}a_c }{3p^2m_*a_{\rm pl}}\right|^{1/3}.
\label{eq:sedef}
\end{align}
Note that 
\begin{equation}
s_e^2=\frac{2}{9p}s.
\label{eq:snue}
\end{equation}

\item{}
As the eccentricity $e\to0$ and the planet mass $m_{\rm pl}\to0$, $n_c =
n$ is the unperturbed mean motion, and the ``exact resonance"
condition $pn_c=(p+1)n_{\rm pl}$ corresponds to $\Delta=0$.   

\item{}
The topology of the phase space determined by this dimensionless
resonant Hamiltonian depends only upon the value of $\Delta$. 

\item{}
The range of the resonant perturbation is $|(p+1)n_{\rm pl}-pn_c|\sim
pn_cs$. This means that
the ``width'' of the resonance is proportional to $m_{\rm pl}^{2/3}$.  
\end{itemize}

\begin{figure*}
   \centering
  \includegraphics[width=\hsize]{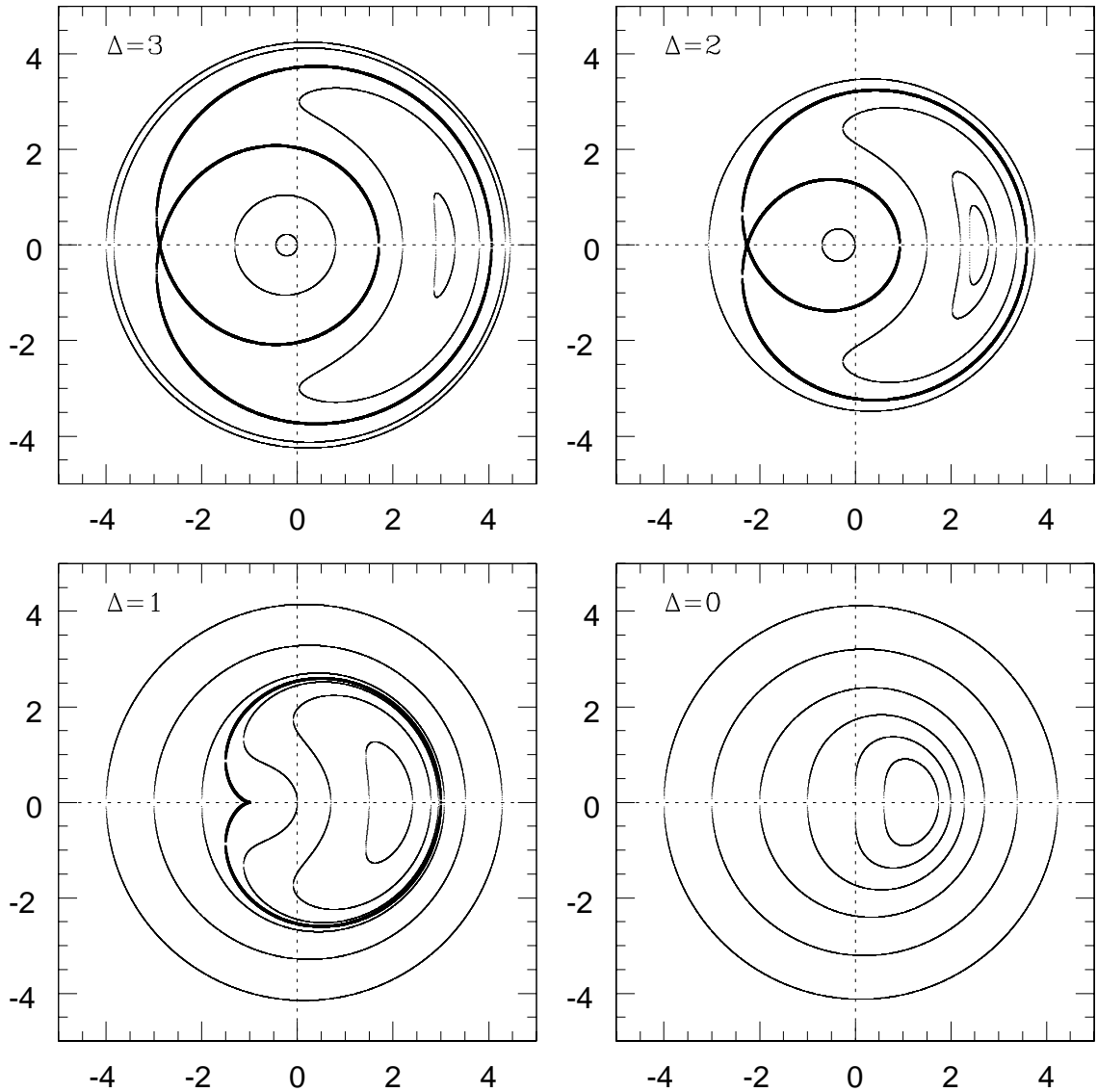}
  \caption{ Level curves of the dimensionless resonance Hamiltonian,
    Equation~(\ref{e:Kres}), for various values of the resonance
    distance $\Delta$, Equation (\ref{eq:delta}). The coordinates are
    $(x,y) = \sqrt{2R}(\cos r,\sin r)$.}
\label{fig:topology}
\end{figure*}   

\begin{figure}
   \centering
   \includegraphics[width=\hsize]{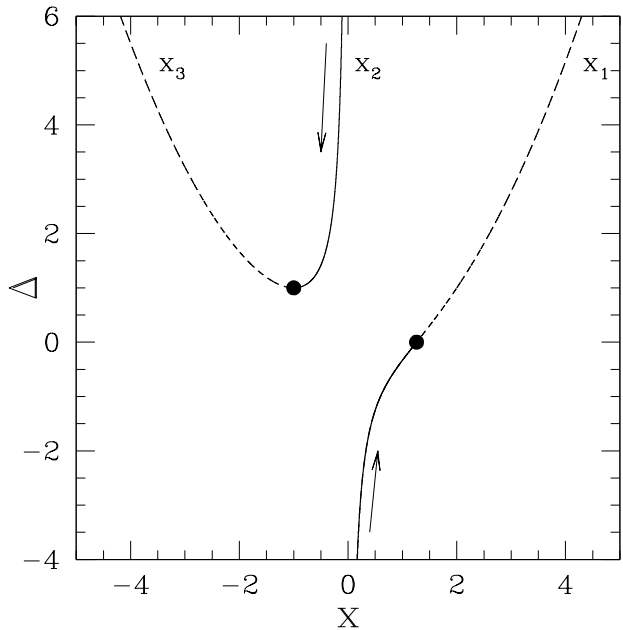} 
   \caption{Fixed points of the  dimensionless Hamiltonian,
     Equation~(\ref{e:Kres}), as a function of the resonance distance
     $\Delta$. The fixed points have $r=0,\pi$ and the variable plotted is
     $x=\sqrt{2R}\cos r$.  
     The arrows indicate the evolution of initially
     circular orbits during the slow growth of the planet, while
     the dashed lines represent fixed points
     that are not populated by test particles during this
     process.   }
\label{fig:fixedpoints}
\end{figure}

The phase-space trajectories follow level curves of the dimensionless
resonant Hamiltonian $K$ (Equation \ref{e:Kres}).
Figure \ref{fig:topology} shows plots of the level curves
for various values of $\Delta$ to illustrate the phase-space
topology.  In these plots, we use the Cartesian variables $(x,y) =
\sqrt{2R}(\cos r,\sin r)$, which are also canonical
($x$ is the momentum and $y$ is the coordinate).  Thus, the origin in these plots
corresponds to zero eccentricity, and the distance from the origin is
$e/s_e$.  The phase-space structure is 
simple when $|\Delta|\gg1$: the trajectories are nearly circles
centered close to the origin.  For $\Delta <1$, there is only one
fixed point and no homoclinic trajectory, but for $\Delta >1$ there
are three fixed points and a homoclinic trajectory exists.  All the
fixed points are on the $x$-axis; they are given by the solutions of
$\partial K/\partial x=0$ which are the real roots of
the cubic equation
\begin{equation}
x^3-3\Delta x-2=0.
\label{e:cubic}\end{equation}
Figure \ref{fig:fixedpoints} plots the locations of the real roots as a
function of $\Delta$. We shall be using only the branches shown by
solid lines: for $\Delta\le0$ 
\begin{equation}
 x_1(\Delta)=(1+\sqrt{1-\Delta^3})^{1/3}+\Delta(1+\sqrt{1-\Delta^3})^{-1/3}, 
\label{eq:roota}
\end{equation}
and for $\Delta\ge0$
\begin{equation}
 x_2(\Delta)=-2\sqrt{\Delta}\cos[(\theta-2\pi)/3], \quad \theta\equiv \cos^{-1}(-\Delta^{-1/3}).
\label{eq:root}
\end{equation}
The level surface of the Hamiltonian that passes through the $x_3$
homoclinic point is a classical quartic curve called the lima\c con of
Pascal. 

\section{Evolution of initially circular orbits if the planet mass
  grows quickly}
  \label{sect:sudden}

\noindent
In this section we determine the distribution of test-particle orbits
that arises if the planet appears quickly. Here ``quickly'' means  over a
timescale long compared to the orbital time but short compared to the
inverse of the frequency scale $sn_c$ (Equation \ref{eq:sdef}); for
reference, at the 2\,:\,1 resonance of Jupiter $(sn_c)^{-1}\simeq
100\yr$. We recognize that this is probably an unrealistic
model of how most planets form, but we
present it as a foil to the slow planetary growth described in
the following section. 

Assume that the planet mass grows suddenly from 0 to $m_{\rm pl}$ at time
$\tau=0$. Since the test particles have zero eccentricity at this
instant, $R=0$ so the resonance Hamiltonian $K=0$ and the fast action
$\Psi=\Lambda=(\mu a_i)^{1/2}$ where $a_i$ is the initial semi-major
axis. Then $n_c$ is a constant of the motion that is equal to
$n_i=(\mu/a_i^3)^{1/2}$, the mean motion just before $\tau=0$ (Equation
\ref{eq:stardef}). The resonant mean motion is $n_{\rm
  res}=(p+1)n_{\rm pl}/p$ and the dimensionless resonance distance (Equation
\ref{eq:delta}) is
\begin{equation}
\Delta=\frac{n_{\rm res}-n_i}{sn_i}.
\label{eq:deldef}
\end{equation}
Thus a uniform initial distribution in mean motion implies a uniform
distribution in $\Delta$.

Observers more commonly work with the period ratio ${\cal P}$, defined
to be greater than unity, i.e., ${\cal P}=n/n_{\rm pl}$ for interior
resonances and $n_{\rm pl}/n$ for exterior ones. Then 
\begin{align}
\Delta=&\frac{1}{s}\left(\frac{{\cal P}_{\rm res}}{{\cal P}_i}-1\right),
\quad {\cal P}_{\rm res}=(p+1)/p,\ p>0\nonumber \\
          =&\frac{1}{s}\left(\frac{{\cal P}_i}{{\cal P}_{\rm res}}-1\right),
\quad {\cal P}_{\rm res}=p/(p+1),\ p<-1.
\end{align}
For $|s|\ll1$, the effects of the resonance are only important 
if $|{\cal P}/{\cal P}_{\rm res}-1|\ll1$ and in this case we can combine
the two formulae above as 
\begin{equation}
\Delta=\frac{1}{|s|}\left(1-\frac{{\cal P}_i}{{\cal P}_{\rm
      res}}\right).
\label{eq:pdef}
\end{equation}

For $\tau>0$ the particles evolve at fixed energy $K$ and action
$\Psi$. For initially circular orbits, $e_i=0$, $\Psi = \sqrt{\mu
  a_i}$, and the constancy of $\Psi$ implies that $a
[1+p-p(1-e^2)^{1/2}]^2 = a_i$ and
\begin{equation}
n = n_i[1+p-p\sqrt{1-e^2}]^3 \simeq n_i(1+\ffrac{3}{2}pe^2).
\label{e:nf}
\end{equation}
Thus, the final mean motion or semi-major axis is determined by the
final eccentricity $e_f$, and our problem is reduced to determining
the distribution of eccentricities $e$ excited by the planet.  Using
Equation (\ref{eq:snue}), Equation (\ref{e:nf}) can be simplified to
\begin{equation}
n=n_i\big(1+\ffrac{2}{3}Rs\big).
\label{eq:final}
\end{equation}

\begin{figure}
   \centering
\includegraphics[width=\hsize]{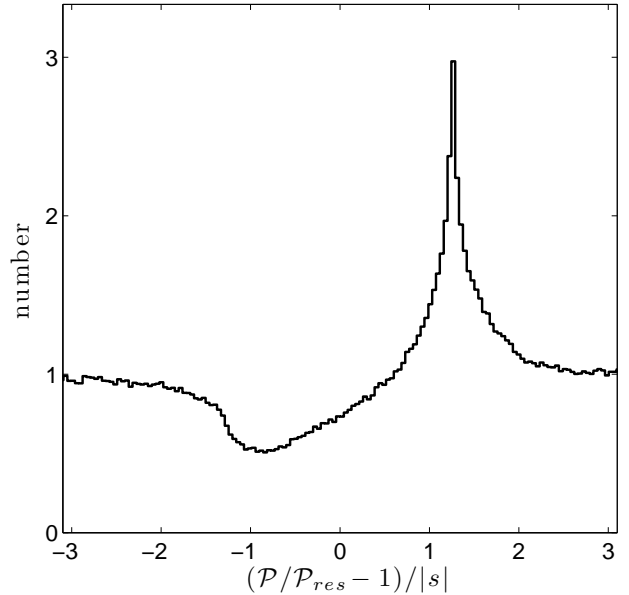}
 \caption{Distribution of period ratios in the vicinity of a
     resonance when the planet mass grows quickly, as described in
     \S\ref{sect:sudden}. Period ratios are
     plotted in units of the resonance strength $|s|$ (Equation
     \ref{eq:sdef}).}
\label{fig:sudden}
\end{figure}   

Now suppose a particle initially has mean motion separated from the
resonance by a fractional distance $fs$ where $f$ is of order unity,
i.e., $n_i-n_{\rm res}=fsn_i$. Then from Equation (\ref{eq:deldef}),
$\Delta=-f$ so the resonance Hamiltonian $K$ (eq.\ \ref{e:Kres}) is
completely specified. The initial conditions are $x=\sqrt{2R}\cos
r=0$, $y=\sqrt{2R}\sin r=0$, and we integrate the equations of motion
$d x/d\tau=-\partial K/\partial y$, $d y/d\tau=\partial K/\partial x$
to the present time; then the present value of $R$ determines the
present mean motion in units of $s$ through Equation (\ref{eq:final}).
Thus the steady-state distribution of mean motions is the same for all
resonances of this type, independent of the planet mass, the resonance
integer $p$, whether the resonance is interior or exterior, etc., so
long as the mean motions are expressed in terms of the dimensionless
parameter $s$ of Equation (\ref{eq:sdef}).

Using this procedure we can compute the expected distribution of mean
motions near a resonance in the limiting case where the planet mass
grows fast. This has been done in Figure \ref{fig:sudden}, where
we have chosen $10^7$ particles with zero initial eccentricity and
initial mean motion chosen randomly in the interval $|{\cal P}_i/{\cal
  P}_{\rm res}-1|\le 3|s|$. Each particle has been followed for a time  chosen
uniformly random between 50 and 150. 

We observe from Figure \ref{fig:sudden} that the distribution of
particles exhibits a peak at period ratios larger than resonance, and
a trough at period ratios smaller than resonance (this
  statement holds for both interior and exterior resonances). 
 The shape resembles a P-Cygni profile in the context of
  spectroscopic emission lines. Using this analogy we can characterize
  the redistribution of systems close to resonances by computing the
  equivalent widths (EWs) in these regions as
\begin{equation}
\mbox{EW}_\pm=\int_{{\cal P}{>\atop <}{\cal P}_{\rm
    res}}\left[\frac{y_f({\cal P})}{y_i}-1\right]\,d{\cal P},
\label{eq:ewdef}
\end{equation}
where $y_i$ is the initial number of particles per unit period ratio 
(assumed uniform) and $y_f({\cal P})$ is the final distribution. Here, the initial
distribution defines our background or ``continuum''. In our
convention a trough
(``absorption line'') has negative EW and a peak (``emission line'')
has positive EW. If particles are rearranged near the
resonance but do not migrate or escape the sum of the EWs
should be zero. The results from the simulation in Figure\
\ref{fig:sudden} give 
\begin{equation}
\mbox{EW}_+=-\mbox{EW}_-=0.685 |s|{\cal P}_{\rm res}.
\label{eq:sudden}
\end{equation}

\section{Evolution of initially circular orbits if the planet mass
  grows slowly}
  \label{sect:adiabatic}

\noindent
We now examine how initially circular orbits of test particles in the
vicinity of resonance evolve as the planet mass grows gradually from
small values.  The slowly varying parameter is the resonance distance $\Delta\propto
m_{\rm pl}^{-2/3}$.  For very small values of planet mass, $|\Delta|$ is
large [for all but the measure zero case in which $(p+1)n_{\rm pl}=pn_c$].
As the planet mass grows to its final value, $|\Delta|$ decreases. 

\subsection{Analysis} 

\noindent
Provided that there is no separatrix-crossing event, an initially
circular orbit of a test particle will evolve with two adiabatic
invariants: $\Psi=\sqrt{\mu a}(1+p-p\sqrt{1-e^2})$ and $A=\oint Rdr =
\oint ydx$, the area enclosed by the phase trajectory in the $(x,y)$
plane.  If a separatrix crossing occurs, there is a discontinuous
change in $A$; this does occur for a range of parameters, as
detailed below.

In the discussion below, we use subscripts $i$ and $f$ to denote
initial and final values of parameters.
For initially circular orbits, the adiabatic invariant $\Psi=\sqrt{\mu a_i}$
 and the constant $n_c$ (Equation \ref{eq:stardef}) is equal to the initial 
 mean motion $n_i$.

Consider first the implications of the adiabatic invariant $\Psi$.
The relations (\ref{eq:deldef}) and (\ref{eq:final}) continue to hold,
except that now the resonance strength $s$ is growing slowly with time. 
Thus, the final mean motion or semi-major axis is determined by the
final eccentricity $e_f$ or its scaled version $R_f$, and our problem is reduced to
determining $R_f$.  We now show that it is possible to determine $R_f$ from the
adiabatic invariance of $A$ for most (but not all) initially circular
orbits.

For initially circular orbits, $A_i=0$. Therefore, these orbits will
adiabatically follow the evolution of the fixed point
close to the origin.   There are two cases to consider, one for each
``side'' of the resonance: 

\paragraph{Test particle is initially wide of the resonance} This
means that the test particle is further from the planet than the
resonance, that is, for an interior resonance ($p>0$) the test
particle is inside the resonance and $n_i > (1+1/p)n_{\rm pl}$, while for an
exterior resonance ($p<-1$) the test particle is outside the resonance
and $n_i < (1+1/p)n_{\rm pl}$. For both cases $\Delta_i$ is negative, and
large in absolute value. Initially circular orbits are near the fixed
point $x_1$ in Figure \ref{fig:fixedpoints}.  As $\Delta$ increases,
these orbits evolve along the fixed point $x_1$, i.e., upward along
the curve labeled $x_1$ in the fourth quadrant of
Figure \ref{fig:fixedpoints}, as indicated by the upward pointing 
arrow.
No matter how large the planet mass grows, $\Delta$ remains negative, 
so the test particle can never evolve onto the dotted portion of the $x_1$
curve.  (The point marked with a black circle at $\Delta=0, x_1=2^{1/3}$
corresponds to the measure-zero case in which $n_i = (1+1/p)n_{\rm pl}$.)
Thus, the final orbits have a fixed eccentricity (Equation \ref{eq:roota}), 
\begin{equation}
e_f=s_ex_1(\Delta_f)\quad\mbox{or}\quad R_f=\half x_1^2(\Delta_f),
\label{eq:efinala}
\end{equation}
and a final mean motion given by Equation~(\ref{eq:final}).

\paragraph{Test particle is initially narrow of the resonance} 

\noindent 
In this case $\Delta_i $ is large and positive and 
initially circular orbits are near the fixed point $x_2$ in Figure
\ref{fig:fixedpoints}.   
As $\Delta$ decreases, these orbits evolve along the fixed point $x_2$,
i.e., downward along the inner curve in the second quadrant in
Figure \ref{fig:fixedpoints}, as indicated by the downward pointing 
arrow.  

No matter how large the planet mass grows, $\Delta$ remains
positive. This branch represents the final orbits for $\Delta_f>1$
(equivalently, $pn_i< (p+1)n_{\rm pl} -psn_c$), and for these orbits the
final eccentricity is fixed at (Equation \ref{eq:root})
\begin{equation}
e_f=s_ex_2(\Delta_f)\quad\mbox{or}\quad R_f=\half x_2^2(\Delta_f).
\label{eq:efinalb}
\end{equation}

A complication in this case is that the fixed point $x_2$ vanishes at
$\Delta=1$, i.e., for $pn_i=(p+1)n_{\rm pl} -psn_c$ where $s$ is the
resonance strength defined by (\ref{eq:sdef}).  For $0<\Delta_f<1$
[i.e., $(p+1)n_{\rm pl}-psn_c< pn_i <(p+1)n_{\rm pl}$], the adiabatic
evolution leads the trajectory to coincide with the separatrix when
$\Delta=1$. There ensues a discontinuous increase in the value of $A$,
which jumps from $A=A_i=0$ to $A=6\pi$
(which is the area enclosed by the separatrix at $\Delta=1$). After
this jump $A$ is again an adiabatic invariant so its final value is
$A_f=6\pi$.  The final orbits in these cases are not described by a
stationary value of the eccentricity, because the trajectory in the
$(x,y)$ plane is neither a fixed point nor a circle centered at the
origin.  Similarly, the final mean motion is not fixed. An approximate
estimate of the mean final eccentricity is 
\begin{equation}
\langle e\rangle_f =s_e\sqrt{A_f/\pi}=\sqrt{6}s_e,
\label{eq:ave}
\end{equation}
but to determine the final distribution of eccentricities and mean
motions in this region it is simpler to integrate the equations of
motion in the resonance Hamiltonian $K$ (see the following
subsection).

\begin{figure}
   \centering
\includegraphics[width=\hsize]{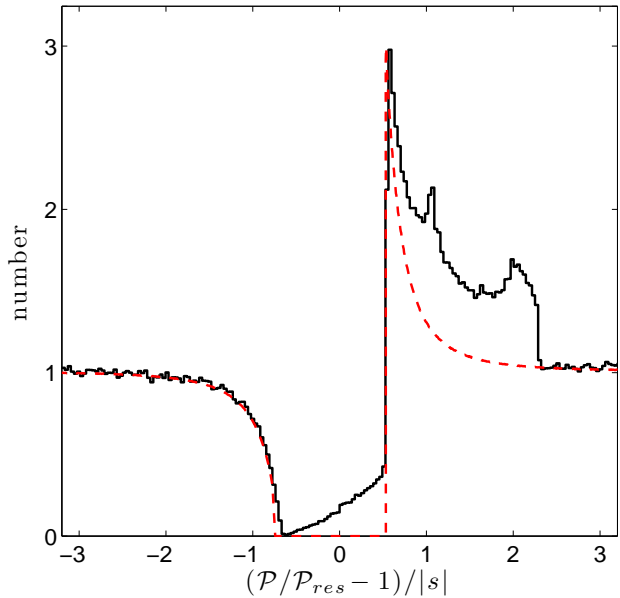}
  \caption{ Distribution of period ratios in the vicinity of a
     resonance when the planet mass grows 
     slowly, as described in \S\ref{sect:adiabatic}. Period ratios are
     plotted in units of the resonance strength $|s|$ (Equation
     \ref{eq:sdef}). The red dashed curves show the expected
     distribution for particles that do not cross the separatrix, as
     given by Equations (\ref{eq:nfni}) and (\ref{eq:nfnidist}). } 
\vspace{20pt}
\label{fig:slow}
\end{figure}

Combining Equations (\ref{eq:final}), (\ref{eq:efinala}), and
(\ref{eq:efinalb}), the final mean motion of particles that do not
cross the separatrix is
\begin{align}
n=&n_i\big(1+\ffrac{1}{3}sx_1^2[(n_{\rm res}-n_i)/sn_i]\big), \quad
p(n_{\rm res}-n_i)>0, \nonumber \\
=&n_i\big(1+\ffrac{1}{3}sx_2^2[(n_{\rm res}-n_i)/sn_i]\big), \nonumber
\\ &\qquad\qquad\qquad\qquad\qquad
p[n_{\rm res}-(1+s)n_i]>0, 
\label{eq:nfni}
\end{align}
where $n_i$ is the initial mean motion, and the characteristic
resonance strength $s$ (Equation \ref{eq:sdef}) is evaluated for the final
planet mass $m_{\rm pl}$. If the initial distribution in mean motion is
uniform, the final density of the mean-motion distribution is given by
\begin{equation}
dN(n) \propto dn_i = \frac{dn_i}{dn}dn,
\label{eq:nfnidist}
\end{equation}
which is easily evaluated numerically from Equation
(\ref{eq:nfni})---see Figure \ref{fig:slow}.  

For these particles there is a gap in the
distribution of final mean motions,
\begin{align}
n_{\rm res}(1-\ffrac{2}{3}|s|) < n& < n_{\rm
  res}(1+\ffrac{2^{2/3}}{3}|s|), \quad p,s>0, \nonumber \\
n_{\rm   res}(1-\ffrac{2^{2/3}}{3}|s|) < n& <   n_{\rm
  res}(1+\ffrac{2}{3}|s|) \quad p,s<0.
\label{eq:gap}
\end{align}
Note that the gap is not symmetric about the exact resonance value,
$n_{\rm res}=(p+1)n_{\rm pl}/p$. The size of the gap is
\begin{equation}
\Delta n_{\rm gap} = 2.587\,(p+1)\Bigg|{m_{\rm pl}\over m_*}{a_cf_p\over a_{\rm pl}p^2}\Bigg|^{2\over3}n_{\rm pl}.
\label{e:ngap}
\end{equation}
The particles that cross the separatrix have librating eccentricities and
mean motions; as we show below, these particles partially fill the gap
and also broaden the peak wide of resonance.

\subsection{Numerics}
\label{sec:slow_num}

\noindent
For our numerical experiments, we assume that the planet mass varies as
\begin{equation}
m_{\rm pl}(t)=m_{p,f}\tanh t/t_{\rm pl}.
\label{eq:tanh}
\end{equation}
The timescale $t_{\rm pl}$ can be thought of as the
formation time of the planet. In simulations with $t_{\rm pl}=0$ the planet
forms suddenly, i.e., the integration is started with the final planet
mass $m_{\rm pl}$. 

We follow the evolution induced by the resonance Hamiltonian $K$ for a
large number of test particles on initially circular orbits, uniformly
distributed in mean motion. We start the integrations at a time when
the planet mass $m_{{\rm pl},i}=10^{-6}m_{{\rm pl},f}$, which means that the initial
characteristic frequency $s_i=10^{-4}s_f$. Thus for any
particle the final resonance distance $\Delta_f=10^{-4}\Delta_i$; we
shall follow particles with $|\Delta_f|\le 3$ which implies that the
initial distribution should be chosen uniform in $\Delta_i$ between
$\pm 3\times10^4$. We use a growth time $t_{\rm pl}=100$ and follow the
particles for a time chosen uniformly random between 250 and
750; we have checked that the results are insensitive to these
choices. The resulting distribution of mean motions is shown in
Figure \ref{fig:slow}; as expected, the distribution mostly
agrees with the distribution derived analytically in the preceding
subsection, but the particles that have crossed the separatrix
partially fill in the resonance gap (Equation \ref{eq:gap}) and enhance the
peak to the right of it. 

The equivalent widths are 
\begin{equation}
\mbox{EW}_+=-\mbox{EW}_-=0.956 |s|{\cal P}_{\rm res},
\label{eq:slow}
\end{equation}
almost $50\%$ larger than the result when the planet grows fast
(Equation \ref{eq:sudden}).

\section{Numerical integrations of the restricted three-body problem}
\label{sec:sims}

\noindent
We ran numerical experiments to follow the motion of test particles
subject to gravitational forces from the central star and an
orbiting massive planet.  We work in the astrocentric reference frame,
in which the equations of motion for a test particle 
can be written as 
\ba
\frac{1}{n_{\rm pl}^2}\frac{d^2\vec{x}}{dt^2}&=&-\mu_1\frac{\vec{x}}{|\vec{x}|^3
} -\mu_2\left[\frac{\vec{x}-\vec{x}_{\rm pl}(t)}{|\vec{x}-\vec{x}_{\rm pl}(t) |^3 }
+\frac{\vec{x}_{\rm pl}(t)}{|\vec{x}_{\rm pl}(t) |^3 }\right],
\ea 
where $\mu_1 =m_{*}/(m_{*}+m_{\rm pl})$ and $\mu_2=1-\mu_1=m_{\rm pl}/(m_*+m_{\rm pl})$. As
usual $a_{\rm pl}$ and $n_{\rm pl}^2 =G (m_*+m_{\rm pl})/a_{\rm pl}^3$ are the semi-major axis and
squared orbital frequency of the planet. The positions of the test
particle $\vec{x}$ and the massive planet $\vec{x}_{\rm pl}$ are normalized
by $a_{\rm pl}$. Times are given in planet years, $2\pi/n_{\rm pl}$.

In these simulations the planet mass $m_{\rm pl}$ or $\mu_2$ is initially
zero and grows during the simulation according to the formula
(\ref{eq:tanh}). 
During this growth the planetary semi-major axis $a_{\rm pl}$ is kept
constant, although there are other plausible choices, e.g.,
$a_{\rm pl}\propto (m_*+m_{\rm pl})^{-1}$, as would be expected if the planet gained
mass isotropically.

If the planet eccentricity $e_{\rm pl}$ is zero and the growth of the planet
mass is sufficiently fast or slow, the results of these simulations 
should be directly comparable to the analytic results we obtained in
\S\ref{sect:sudden} and  \S\ref{sect:adiabatic}. 

\subsection{Initial conditions}

\label{sec:ic} 

\noindent
We typically consider planet masses $m_{\rm pl}$ in the range 
$10^{-4}$--$10^{-3}m_*$, or 0.1--1 Jupiter mass for a solar-mass host
star. For comparison most  {\it  Kepler} planets have masses between 0.01 and
0.1 Jupiter masses, and most planets discovered by radial-velocity
measurements have masses between 0.1 and 10 Jupiter masses. 

We assume for simplicity that the test particle is inside the massive
planet, i.e., we consider interior resonances only. This is a
plausible simplification: inner planets are usually smaller than
outer planets in the {\it  Kepler} and radial-velocity samples because
planets with smaller semi-major axes are easier to detect. 

The semi-major axes $a$ of the test particle and planet are determined
using a fitting function for the probability distribution of semi-major axes
in the {\it  Kepler} sample, after accounting for geometric selection effects
\citep{TD11}: 
\ba 
dp(a)=0.656 \frac{(a/a_0)^{3.1}} {1+(a/a_0)^{3.6}}
\frac{da}{a}, \quad a<1.15\,\mbox{AU}
\label{eq:epsilon}
\ea where $a_0=0.085\,$AU.  We then generate an initial distribution
in the period ratio $\mathcal{P}$ ($\mathcal{P}>1$) of a two-planet
system by generating two random variables, $a_1$ and $a_2$, from this
probability distribution and computing
$\mathcal{P}=\max\left\{(a_1/a_2)^{3/2}, (a_2/a_1)^{3/2}\right\}$.
This procedure generates a smooth initial distribution in period
ratio. For the numerical integrations, we set the semi-major axis of
the massive planet to unity and the semi-major axis of the test
particle is then $\mathcal{P}^{-2/3}$.

In some simulations the test particles have non-zero eccentricities $e$
and/or inclinations $i$. These are assumed to be randomly distributed following a Rayleigh law, 
\ba
dp=\frac{x\,dx}{\sigma_x^2} \exp(-\ffrac{1}{2} x^2/\sigma_x^2),
\label{eq:sigma_e}
\ea
where $x=e$ or $i$ and $\sigma_x$ is an input parameter that
is related to the mean and rms eccentricity or inclination by $\langle
x\rangle=\sqrt{\pi/2}\sigma_x=1.253\sigma_x$, $\langle
x^2\rangle^{1/2}=\sqrt{2}\sigma_x=1.414\sigma_x$.

We treat the planet and host star as point particles, i.e., we do not
account for possible collisions of the test particles with either
body. 

\subsection{Numerical results and comparison with analytic theory}

\noindent
We start by considering two simple fiducial models to compare with our
previous theoretical results.
First, we consider a model for a slowly growing planet mass,
which contains $10^4$ test particles and has 
planet-formation timescale $t_{\rm pl}=10^{4}$ $ [2\pi/n_{\rm pl}]$ (or $10^{4}$
orbits of the exterior planet). 
Second, we consider a model for a rapidly growing planet mass,
which contains $2\times10^4$ test particles and starts with the final
planet mass, i.e., $t_{\rm pl}=0$.
Both models have a final planet mass $m_{\rm pl}=0.001m_*$ (1 Jupiter mass),
a planet eccentricity $e_{\rm pl}=0$, and zero eccentricities
and inclinations for the test particles
(i.e., $\sigma_e=\sigma_i=0$ in Equation \ref{eq:sigma_e}).
The initial distribution of semi-major axes or periods of the
test particles is obtained using the procedure in \S\ref{sec:ic}, with
the period ratio restricted to the range $1.3\le \mathcal{P}\le
2.5$. 

In the simulation for the slow case, $\approx 4$\% of the test particles 
were lost to escape orbits, mostly from orbits initially near the 
planet with $\mathcal{P}_i \lesssim  4/3$, while for the fast case
the same happens for $\approx 10$\% of the test particles,
almost all from orbits with $\mathcal{P}_i \lesssim  7/5$.

\begin{figure}[h]
\centering
\includegraphics[width=1.0\hsize, height=0.9\hsize]{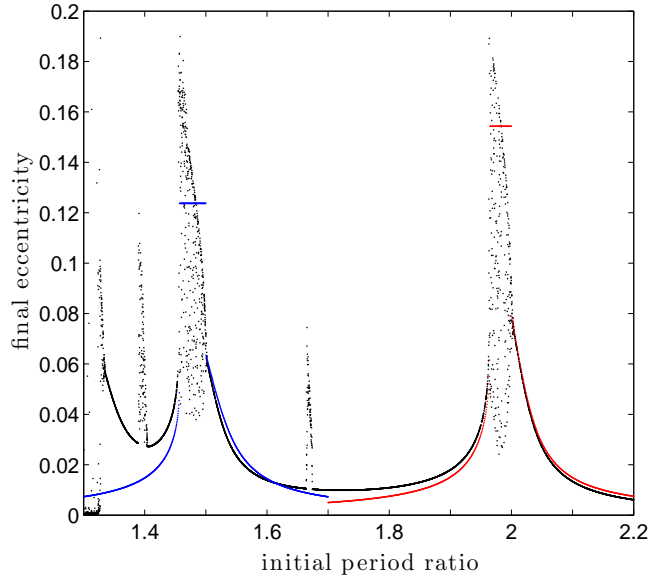}
\caption{Final eccentricity as a function of the initial period ratio
  $n_i/n_{\rm pl}$ for a simulation with $10^4$ test particles and parameters
  $m_{\rm pl}=0.001m_*$, $t_{\rm pl}=10^{4}$, $e_{\rm pl}=0$, and $\sigma_e=0$. Results  are shown after $5\times10^4$ planet orbits.  
  The blue and red lines
  indicate the analytic results for a slowly growing planet, for the 3\,:\,2 and
  2\,:\,1 resonances, respectively. The analytic results are only shown for
  particles that do not cross the resonance separatrix, since particles
  that do cross have librating eccentricities (these produce the elongated
  ovals filled with scattered points). The horizontal lines denote the
  approximate analytic prediction for the mean eccentricity of
  particles with oscillating eccentricity (Equation [\ref{eq:ave}]).  }
\label{fig:e_f}
\end{figure}

\subsubsection{Final eccentricity}

\noindent
In Figure \ref{fig:e_f}, we compare our numerical results for the
final eccentricity as a function of the initial period ratio with
those obtained from the analytic formalism of \S\ref{sect:adiabatic}
for a slowly growing mass planet.
As expected, substantial eccentricities are excited in orbits that
start close to the first-order (4\,:\,3, 3\,:\,2 and 2\,:\,1) resonances. The
eccentricities are also excited in smaller ranges around the
second-order (7\,:\,5 and 5\,:\,3) resonances, a result that is not captured
by our simple analytic theory. Analytic curves for the 3\,:\,2 and 2\,:\,1
resonances (blue and red, respectively, from Equations \ref{eq:nfni} and
\ref{eq:nfnidist}) are plotted for orbits that do not cross the
resonant separatrix, since these have stationary forced eccentricity
(compare the red dashed lines in Figure \ref{fig:slow}). The analytic
curves agree quite well with the simulations close to the resonances.
The deviations grow as we move away from the resonances, since the
approximation of a single resonance becomes less and less accurate.
For orbits that do cross the separatrix, the figure shows an approximate estimate for the mean eccentricity from Equation (\ref{eq:ave}) as a horizontal
line; the extent of this line marks the range of period ratios
corresponding to such orbits, from Equation (\ref{eq:gap}). 
In this snapshot of the simulation, the regions in which the forced 
eccentricity oscillates are marked by clouds of points; the widths 
of these clouds agree reasonably well with the analytic theory, 
although the theoretical estimates of the mean eccentricities, marked
by horizontal red and blue bars, are somewhat too high.

For the case when the planet mass grows quickly the 
final eccentricity is never stationary: it
oscillates between zero and a maximum value similar to that
in Figure \ref{fig:e_f}.
The results obtained from 
the resonance Hamiltonian integration in 
\S \ref{sect:sudden} agree quite well with the numerical 
simulations. Near the 2\,:\,1 and 3\,:\,2 resonances, both produce  
a similar cloud of  final eccentricities 
for a given initial period ratio.

\begin{figure}[h]
\centering
\includegraphics[width=8.5cm]{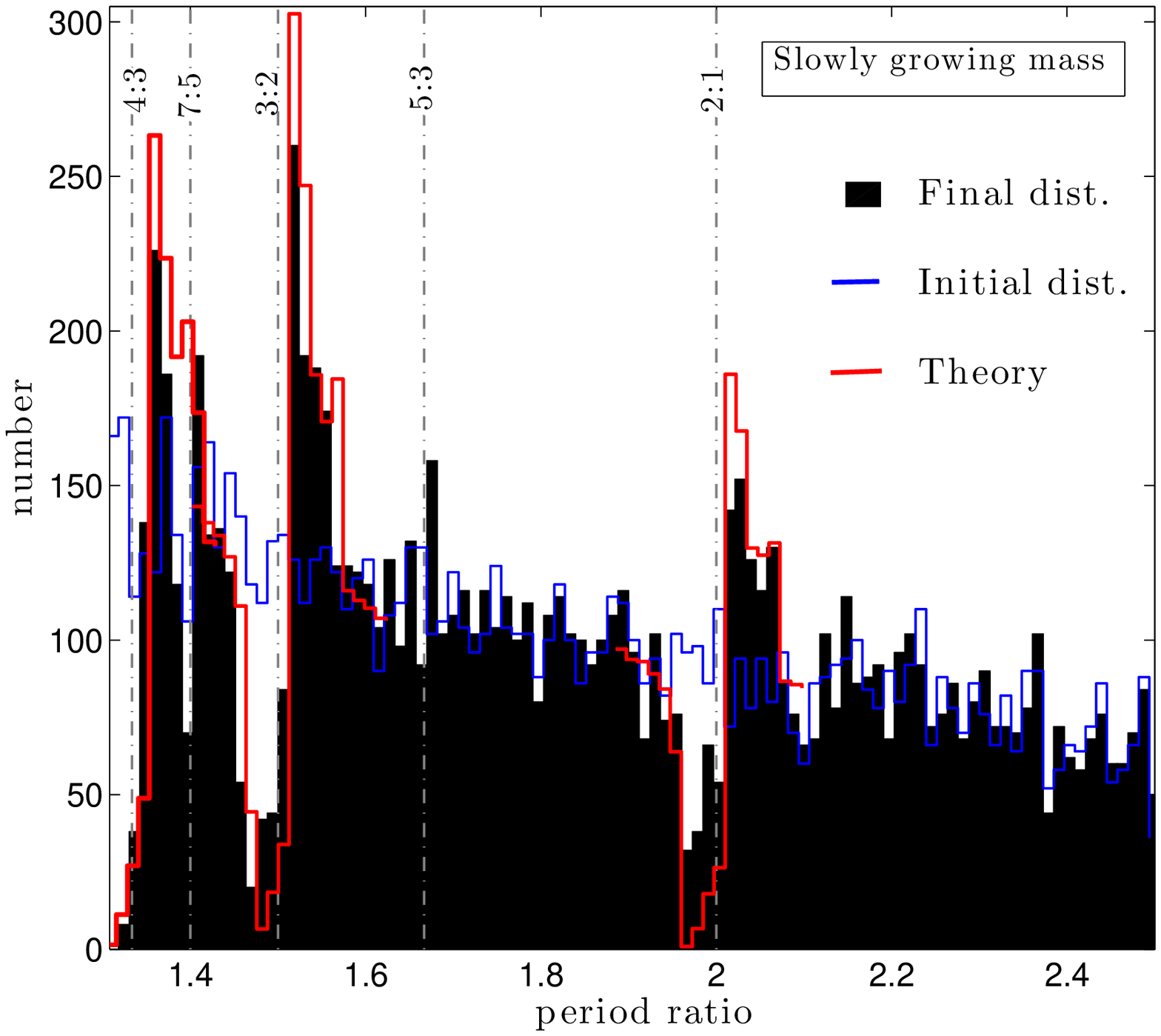}
\includegraphics[width=8.6cm]{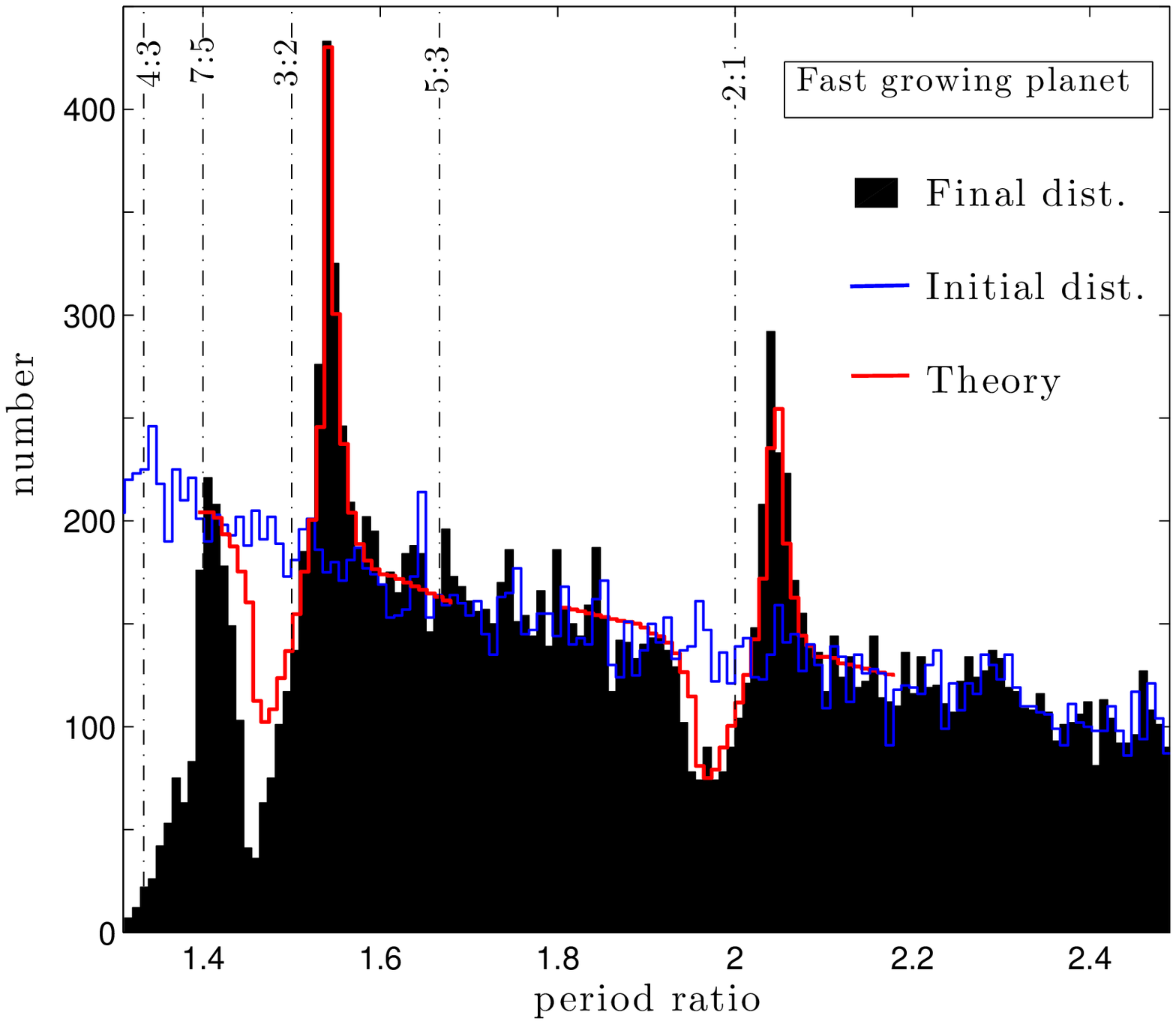}
\caption{Number density of test particles as a function of the period ratio
  $\mathcal{P}$ (black histograms), 
  for the fiducial three-body integrations with parameters
  $m_{\rm pl}=0.001m_*$, $e_{\rm pl}=0$,  $\sigma_e=0$, and $\sigma_i=0$. The 
  initial distribution in period ratio is
  given by Equation  (\ref{eq:epsilon}) and shown by blue lines.
  The upper panel shows the distribution after 
  $5\times10^{4}$  planet  years for a simulation 
  with growth time $t_{\rm pl}=10^4$ planet years, as well as the 
  theoretical PDF around the 4\,:\,3, 3\,:\,2, and 2\,:\,1 resonances,
  shown in red lines. The latter distribution is 
  obtained as in \S\ref{sec:slow_num} (see Figure \ref{fig:slow}) and
  rescaled to the number of systems in the three-body integrations.
 The lower panel shows the same for a set of three-body
 integrations in which the growth time $t_{\rm pl}=0$; in this panel
 the theoretical PDF is obtained as in  \S\ref{sect:sudden}
 (see Figure  \ref{fig:sudden}).\\
 }
\label{fig:dist_fiducial}
\end{figure}

\subsubsection{Period-ratio distribution}

\noindent
In Figure \ref{fig:dist_fiducial} we show the distribution in period
ratio for the test particles in the fiducial simulations. The upper 
and lower panels show the results for the slowly and fast
growing mass cases, respectively. 
We also plot in red lines the probability density function (PDF) 
around the first-order resonances,
as computed in \S\ref{sect:sudden} and \S\ref{sec:slow_num} 
respectively (see black lines in figure \ref{fig:sudden} and
\ref{fig:slow}), scaling our results to the mean
number of initial planets in the simulation as expected
from the background distribution in Equation (\ref{eq:epsilon}). 

As predicted by the analysis for a slowly growing planet in
\S\ref{sect:adiabatic}, there is a peak in the number of planetary
systems at period ratios slightly larger than the exact resonances and
a deficit at period ratios smaller than the resonance position.  The
strongest enhancements are at the first-order 4\,:\,3, 3\,:\,2, and
2\,:\,1 resonances, but the effect is also weakly visible at the
second-order 7\,:\,5 and 5\,:\,3 resonances.  Note, however, that the
7\,:\,5 resonance is close enough to 4\,:\,3 that the assumption of an
isolated resonance used in the theory is suspect. 
Similarly, the 4\,:\,3 resonance is close enough to
the planet that a significant fraction of the particles at smaller
period ratios are ejected by the planet.

The features seen in the three-body integrations at the first-order
resonances are roughly consistent with the predictions of
\S\ref{sect:adiabatic} (red lines in Figure
\ref{fig:dist_fiducial})). Specifically, the theory is able to
reproduce reasonably well the position and height of the peaks and the
width and depth of the gaps at the 3\,:\,2 and 2\,:\,1 resonances.

Similarly, as seen in the lower panel of Figure \ref{fig:dist_fiducial}, 
 the theory of \S\ref{sect:sudden} for the rapidly growing planet
agrees reasonably well with the simulations around the
3\,:\,2 and 2\,:\,1 resonances. We observe, however, 
that the gap narrow of the 3\,:\,2 resonance is deeper
than the single-resonance theory predicts, which is mainly due to a
fraction of the particles that are excited to high eccentricities
($e\simeq 0.2$--$0.6$)
and larger period ratios (310 particles with initial
period ratio $1.4<\mathcal{P}_i<1.5$ have a final period ratio
$1.6<\mathcal{P}<2.5$, while 16 are ejected from the system).
The agreement with the single-resonance approximation is even worse closer to
the planet: in fact, this approximation breaks down completely when
first-order resonances overlap, which is expected for period ratios 
$<1.33$ for a Jupiter-mass planet \citep{wisdom:1980}. Additional
discrepancies arise from ejection of test particles by the planet,
which occurs for $77\%$ of the particles with $\mathcal{P}_i<7/5$. 

In summary, the single-resonance analysis of \S\ref{sect:sudden}
and \S\ref{sect:adiabatic} is able to reproduce the main features of the
three-body integrations around the 3\,:\,2 and 2\,:\,1 resonances. The
period-ratio distributions for both slow and fast planet growth
exhibit peaks wide of the resonance (period ratios larger than the
resonance value) and troughs or gaps narrow of the resonance. This
result suggests that a broad range of planet-formation processes on
intermediate timescales will produce a similar period-ratio
distribution, so long as the planet does not migrate. However, the
distributions differ in detail when the planet mass is increased
slowly or fast.  Specifically, the slow case produces both peaks and troughs
that are sharper and lie closer to the resonance.

\begin{deluxetable}{ccccc}{*h}
\centering
\tablewidth{0pc}
\tablecaption{Equivalent widths around resonances  }
\tablehead{
Resonance & 
EW$_+$  &
EW$_-$  &
$|$EW$_\pm|$ \\
&&&from Equation (\ref{eq:slow})}
\startdata
4\,:\,3 &  0.0073$^{a}$    &$-0.034$~ &0.033 \\
7\,:\,5 &  0.0043~~    &$-0.0051$ &|  \\
3\,:\,2 &  0.033~~~   &$-0.034$~ & 0.033  \\
5\,:\,3 &  0.0033~~    &$-0.0036$ & |  \\
2\,:\,1 &  0.030~~~ &$-0.030$~ & 0.034
\enddata
\tablecomments{Equivalent widths (Equation \ref{eq:ewdef}) 
from three-body integrations, for  slow growth of the planet mass.}
\tablenotetext{a}{Note that $82 \%$ of the test
particles around 4\,:\,3  are lost to escape orbits and 
this decreases the value of EW$_+$, while 
slightly increasing the absolute value of EW$_-$.}
\label{tab:EW}
\end{deluxetable}

\subsubsection{Equivalent widths around resonances} 

\noindent
The equivalent widths measured from the fiducial simulation 
of a slowly growing planet are shown
in Table \ref{tab:EW}, along with the predictions from Equation
(\ref{eq:slow}). In most cases the EWs on either side of the resonances are almost
equal and opposite, confirming that the resonances lead to shuffling
of the test particle semi-major axes but not any overall loss or 
accumulation.

As expected, the largest EWs are at the 3\,:\,2 and 2\,:\,1
resonances, $\simeq 0.033$ and $\simeq 0.030$, respectively.  The EWs
at 3\,:\,2 and 2\,:\,1 agree  with the single-resonance
approximations (Equation \ref{eq:ewdef}).  The 4\,:\,3 resonance
agrees less well because 182 test particles around this resonance were
ejected.  But even in this case the single-resonance theory matches at
least the width of the trough, EW$_-$.

For the fast mass increase case, we measure 
EW$_-=-0.0262$  and 
EW$_+=0.0254 $ around the 2\,:\,1 resonance,
and the single-resonance theory (Equation \ref{eq:sudden}) yields 
$|$EW$_\pm|=0.0244$, in reasonably good agreement. 
At the 3\,:\,2 resonance, 
EW$_-=-0.037$ and 
EW$_+=0.030$, 
while the single-resonance theory predicts 
$|$EW$_\pm|=0.0237$. 
Here the difference between the single-resonance theory and the 
three-body integrations is mostly due to the excitation of 
somewhat larger eccentricities in the integrations that
tend to populate the peak wide of 3\,:\,2 with more
particles, while excavating a deeper gap around 
this resonance; 310 test particles with $1.4<\mathcal{P}_i<1.5$ are promoted
to orbits with $\mathcal{P}>1.6$ and 16 to escape orbits,
which accounts for the difference in magnitude
between EW$_-$ and EW$_+$.
We have checked that reducing the mass of the 
perturber improves the agreement between the 
single-resonance theory and the integrations.

\begin{figure*}
\centering
\includegraphics[width=16cm, height=7cm]{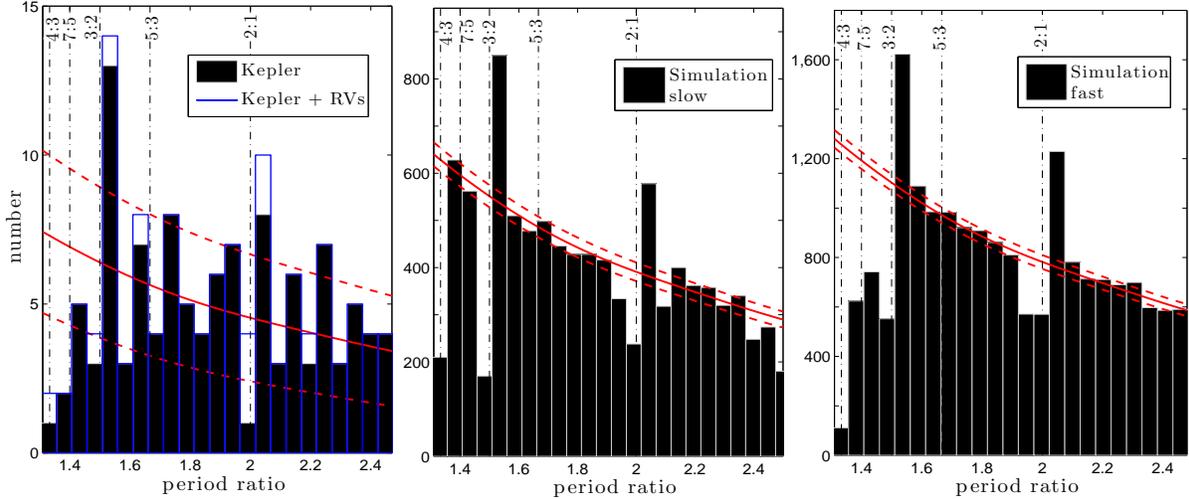}
\caption{Number density of systems as a function of the
period ratio. 
The left panel shows the
{\it Kepler} sample (black histogram)
and the {\it  Kepler} sample plus the RV sample (blue line).
The middle and right panels show the
results of the fiducial simulations for a slowly and rapidly
growing planet, respectively.
The solid red lines indicate the
initial distribution in period ratio given by Equation  
(\ref{eq:epsilon}), scaled to the number of systems
in each panel (ignoring the small RV contribution in the left panel) 
and the red dashed lines its 1--$\sigma$ error bands.
The vertical dot-dashed lines show the positions of the 
first- and second-order resonances.
All histograms have the same binning of $0.05$.\\
\bigskip}
\label{fig:sim+obs}
\end{figure*}

\subsection{Dependence on input parameters}
\noindent 
All of the simulations described so far consider a Jupiter-mass planet
with zero eccentricity and test particles in initially circular and
coplanar orbits, so that a direct comparison with the single-resonance
theory can be made.  Here we comment on how the distributions in
period ratio are affected by different values of the planet mass and
eccentricity, and the initial rms eccentricity and inclination of the
test particles.

\subsubsection{Varying the planet mass $m_{\rm pl}$}
\noindent
We have run sets of three-body integrations in which the mass of the
planet varies over the range $10^{-4}$--$10^{-3}m_*$. These
experiments show that the integration results obey the scalings
predicted by the single-resonance theory: in particular 
the eccentricity of the test particles scales  
as $s_e\propto m_{\rm pl}^{1/3}$, while
the equivalent width varies as $|$EW$_\pm |\propto s \propto m_{\rm pl}^{2/3}$. 
Thus, varying the planet mass simply produces re-scaled versions of the
PDF around the 3\,:\,2 and 2\,:\,1 resonances shown in 
Figure \ref{fig:dist_fiducial}.

These scalings are more difficult to study for resonances that are closer 
to the planet, both because of particle ejections and possible
resonance overlap. Nevertheless, in most cases the analytical
single-resonance theory allows us to scale predictions for the
distribution of period ratios to different resonances and different
planetary masses.

\subsubsection{Varying the eccentricities: $e_{\rm pl}$  and $\sigma_e$}

\noindent
Varying the eccentricities of the  test particles can
strongly modify the final distribution of period ratios.  In
particular, values of $\sigma_e$ (Equation \ref{eq:sigma_e}) 
that are larger than the eccentricity
scale $s_e$ (Equation \ref{eq:sedef}) tend to suppress the 
characteristic peak+trough or
``P-Cygni'' feature around first-order resonances.  For reference,
from Equation (\ref{eq:sedef}) the eccentricity scale for a
Jupiter-mass planet at the 
3\,:\,2 and 2\,:\,1 resonances is 0.05 and 0.063,
respectively. For values of $\sigma_e$ lower than $s_e$ the PDFs are
similar to those starting with initially circular orbits
($\sigma_e=0$).

Non-zero eccentricity of the perturber is expected to introduce
chaotic zones in the vicinity of the resonances, such that 
initially circular orbits are excited to higher eccentricity 
 over a wider range in period ratios compared with the
case of a zero-eccentricity perturber.
In numerical simulations with perturber eccentricities $e_{\rm pl}=0.05$--0.1, 
we observe that a large fraction of test 
particles with $\mathcal{P}_i<3/2$ are ejected to escape 
orbits, leaving a strong gap around 3\,:\,2 and almost no
evidence of a peak wide of the resonance.
Additionally, the gap around the 2:1 resonance is observed to be wider
and less deep than for a zero-eccentricity perturber, while the peak wide of the resonance 
is observed to be somewhat weaker. 
 Non-zero eccentricities could also
lead to slow evolution of the period-ratio distribution on timescales
much longer than the $10^4$ planet orbits used in the fiducial
simulations; we have carried out extended integrations for up to $10^5$
planet orbits but see no additional changes in the period-ratio
distribution. 
In summary, the non-zero eccentricity of the perturber 
tends to smear out the characteristic ``P-Cygni"
profile that we observe in the zero-eccentricity
simulations.

\subsubsection{Varying the inclinations: $\sigma_i$}
\noindent
We have run three-body integrations with parameters $\sigma_e=0.01$
and $\sigma_i=0.1$.  The latter parameter gives a mean inclination of
$\langle i\rangle\approx 7^{\circ}$, a value that is somewhat higher
than estimates of the mean inclinations in {\it  Kepler} multi-planet
systems: $< 5^{\circ}$ \citep{TD11} or $1.0^\circ$--$2.3^\circ$
\citep{FL12}.  Although the dominant features around the first-order
3\,:\,2 and 2\,:\,1 resonances are still present, the ``P-Cygni"
profile shows a less sharp peak that is shifted to larger period ratios, 
 and gaps that are slightly wider than in the zero-inclination case.
More specifically,  we find that at 2\,:\,1
the equivalent widths slightly increase to
  $|$EW$_\pm|=0.033$  
(see values in Table \ref{tab:EW} for comparison with the
 fiducial simulation), while
at 3\,:\,2 we observe that more particles are ejected to
escape orbits relative to the zero-inclination case, resulting in an  
increase in magnitude of 
EW$_-$ to $-0.038 $ and a decrease of EW$_+$ to $0.021$. 
Finally, the gap at the second-order 
7\,:\,5 resonance gets wider by a factor of roughly two.

\begin{figure*}
\centering
\includegraphics[width=16cm, height=7cm]{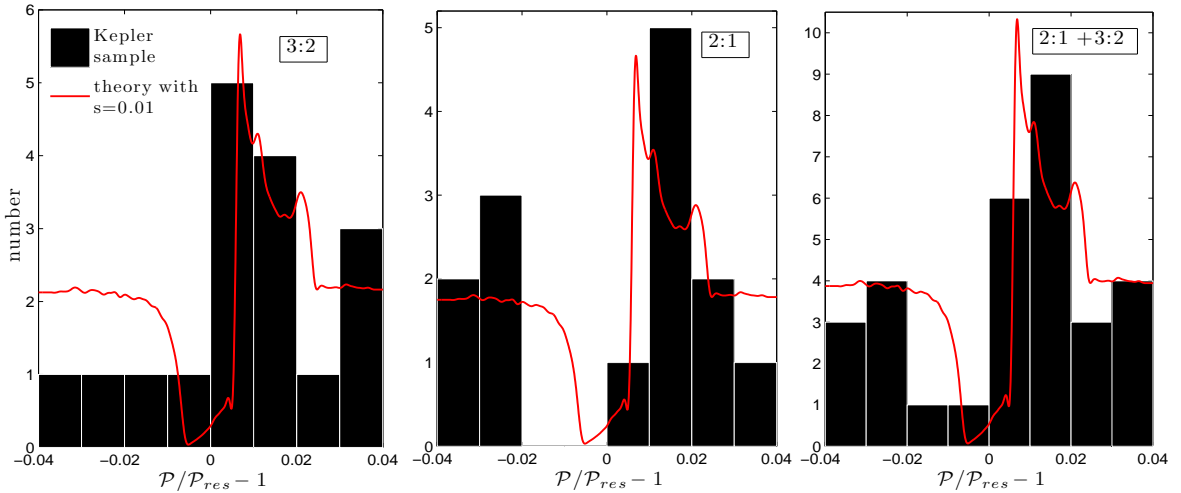}
\caption{Distribution in period ratio of the {\it  Kepler} two-planet systems
  near the 3\,:\,2 and 2\,:\,1 resonances (left and middle panels). In
  the right panel the two distributions are added together (for given
  values of the shifted and normalized period ratio $\mathcal{P}/\mathcal{P}_{\rm
    res}-1$).  The red line indicates the theoretical distribution for a slowly
  growing planet (the same as in Figure \ref{fig:slow}) with a
  strength 
  parameter (Equation \ref{eq:sdef}) $s=0.01$, normalized to the number of {\it  Kepler} planets in
  each panel.
  \\
  \bigskip }
\label{fig:obs_resonances}
\end{figure*}

\section{Comparison with the {\it  Kepler} systems}

\noindent
We now compare our results with the {\it  Kepler} catalog of 242 two-planet
systems as obtained from \citet{FL12}.
We do not consider systems with more than two planets
since we have not investigated three-body resonances in this paper. We
also compare with the 33 two-planet systems discovered by radial
velocity (RV) observations, using the \textit{The Exoplanet Orbit
  Database} as of September 2012 \citep{wright}. 
We limit our analysis to a range in period 
ratio $\mathcal{P}$ of $1.3$--$2.5$, in which the first-order resonances
are 2\,:\,1, 3\,:\,2, and 4\,:\,3. 
After this cut our sample is reduced to 116 and 10 two-planet systems 
from the {\it  Kepler} and RV catalogs, respectively.

In Figure \ref{fig:sim+obs}, we show the distribution of period ratios
for the observations (left panel) and the fiducial simulations.  
We also show the expected
initial distribution of period ratios as determined by the algorithm
described in the paragraph containing Equation (\ref{eq:epsilon}),
along with its 1--$\sigma$ confidence limits to suggest whether 
peaks and dips are significant. 

The observations exhibit statistically significant peaks just outside
(larger $\mathcal{P}$ than) the 3\,:\,2 and 2\,:\,1 resonances,
although the peak at 2\,:\,1 only exceeds 1.5--$\sigma$ significance
when the RV planets are included in the sample (these statements
depend on the binning, which has been chosen to maximize the
significance of the near-resonant features). The observations show
troughs narrow of the 3\,:\,2 and 2\,:\,1 resonance, although these
are only marginally significant (1--$1.5\sigma$).  There is also a
deficit of systems with period ratios $\mathcal{P}\lesssim 1.4$;
this likely arises from the depletion of planets that suffer close
encounters and are scattered onto collision or ejection orbits. There
are no statistically significant features at any second-order
resonance.  Therefore, we will concentrate our analysis on the
first-order 2\,:\,1 and 3\,:\,2 resonances.

\subsection{The 3\,:\,2 and 2\,:\,1 resonances}

\noindent
As pointed out by \citet{LF11} and \citet{FL12} and also shown in
Figure \ref{fig:sim+obs}, two-planet systems close to the 3\,:\,2 and
2\,:\,1 resonances prefer period ratios wide of the resonance within a
few percent; in particular, the data show significant peaks wide of
the resonances and perhaps a trough narrow of the resonance in the
case of the 2\,:\,1 resonance. Here we compare the distribution of
planetary systems around the 3\,:\,2 and 2\,:\,1 resonances with our
simple models to test whether they are consistent.

Hereafter we shall concentrate on the more realistic model
in which the mass grows slowly and will refer to the distribution
obtained using the single-resonance approximation (\S\ref{sect:adiabatic} 
and Figure \ref{fig:slow}) as our theoretical PDF.

In Figure \ref{fig:obs_resonances} we show the distribution of systems
as a function of the shifted and normalized period ratio
$\mathcal{P}/\mathcal{P}_{\rm res}-1$. Positive and negative values of
this variable lie wide or narrow of the resonance, respectively.  For
comparison, we show our theoretical PDF normalized to the
number of planets in each panel.  Here we just consider one value of
the strength, $s=0.01$, which is equivalent to
a planet mass $m_{\rm pl}\simeq0.3 M_J$ or $m_{\rm pl}\simeq 0.4 M_J$ at
3\,:\,2 and 2\,:\,1, respectively. The corresponding equivalent width
is $EW_\pm=\pm0.00956\,\mathcal{P}_{\rm res}$ 
(see Equation \ref{eq:slow}).

We observe in the left panel of Figure \ref{fig:obs_resonances} that
the theoretical PDF produces a peak similar in height and location to
that observed in the {\it  Kepler} sample around the 3\,:\,2 resonance, but
the {\it  Kepler} sample shows no sign of the expected gap. In contrast, in the middle
panel the gap inside the 2\,:\,1 resonance seems to be larger in the
{\it Kepler} observations than in the theoretical PDF.  Finally, in the
right panel where we add the 3\,:\,2 and 2\,:\,1 resonances together,
the overall shape of the period distribution seems to agree, at least visually,
with that of the theoretical PDF. More quantitative comparison
requires statistical tests, to which we now turn. 

\begin{figure}
\centering
\includegraphics[height=12cm]{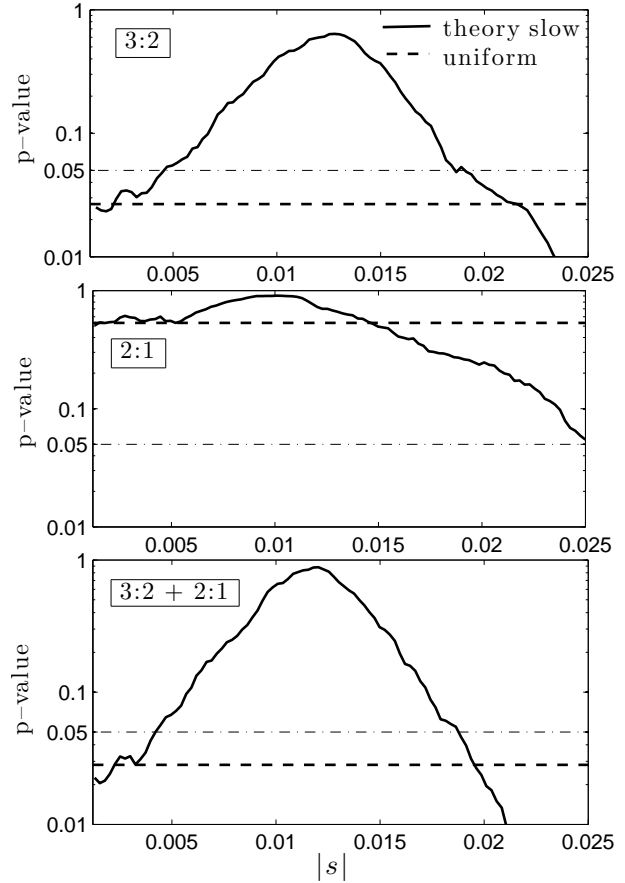}
\caption{$p$--values as a function of the resonance strength $s$
  (Equation \ref{eq:sdef}) obtained from a Kolmogorov--Smirnov test
  comparing the {\it Kepler} two-planet sample with the theoretical
  PDF for a slowly growing planet that does not migrate (solid
  line). The horizontal dashed line represents a uniform density
  distribution, and the dash-dot line represents a significance level
  $p=0.05$---models below this line are excluded at the 95\% confidence
  level.  We limit the test to the range
  $|\mathcal{P}/\mathcal{P}_{\rm res}-1|<0.04$ to maximize its power. }
\label{fig:p_values}
\end{figure}

\subsubsection{K-S test}
\label{sec:ks}
\noindent
We use a Kolmogorov-Smirnov test (K-S test) 
to determine the probability that the observed sample 
is drawn from the same distribution as our
theoretical distribution. 
Since we have found that the PDF obtained from the
single-resonance model is a good approximation to the PDF obtained from 
three-body integrations starting with low-eccentricity,
low-inclination orbits,  we shall use the single-resonance predictions for the
subsequent analysis.

Note that the {\it Kepler} sample contains planets with a wide range of
masses. Thus we are fitting the data to some loosely defined
``typical'' value of the planet mass $m_{\rm pl}$ or dimensionless resonance
strength $s$ (Equation \ref{eq:sdef}).  

In Figure \ref{fig:p_values} we show the $p$--values
obtained from the K-S test for the {\it  Kepler} planets near the 3\,:\,2 and
2\,:\,1 resonances, and for the sum of these two distributions at
given $\mathcal{P}/\mathcal{P}_{\rm res}-1$ (from top to bottom panels). 
For reference, we also show the $p$--value obtained by comparing the
{\it Kepler} planets to a uniform distribution (horizontal dashed line).
In this test, we have restricted the samples to the 
range $|\mathcal{P}/\mathcal{P}_{\rm res}-1|<0.04$. We have experimented
with other ranges between 0.03 and 0.1 and found that 0.04 provides
the most stringent comparisons. For ranges larger than 0.04 there are too many 
particles far from resonance that dilute the signal; also the test
becomes biased by large-scale gradients in the period ratio
distribution. For ranges smaller than 0.03 there are too few planets
left in the sample. 

As usual, if the $p-$value is below a given value $\alpha$
there is a probability of $1-\alpha$ that the {\it  Kepler} 
data are not drawn from the PDF predicted by the single-resonance model. 
Thus, in each panel we add a horizontal dot-dashed 
line at $0.05$ to discriminate between models
at the $95\%$ confidence level.
The middle panel of Figure \ref{fig:obs_resonances} shows that there
is no statistically significant signal at the 2\,:\,1 resonance, in
the sense that a uniform distribution of period ratios is consistent
with the data (at $p=0.53$). This result does not exclude the
possibility that a significant signal would be revealed by some other
more sensitive test. 
In the top and bottom panels, there is a
significant signal in that the uniform distribution is excluded at
about the 97\% confidence level.

These results differ from those of \citet{FL12}, who find significant
evidence that the distribution in period ratios around the 3\,:\,2 and
2\,:\,1 resonances are not drawn from a smooth distribution
($p=0.0046$ and 0.029, respectively), both more than an order of
magnitude smaller that our $p$-values. We attribute this difference
mainly to our restricted data set: we use only two-planet systems
whereas Fabrycky et al.\ use all planet pairs in multi-planet systems,
so our sample is $\sim 3$ times smaller. The difference might also hint that
near-resonant features are more common in multi-planet systems 
than in two-planet systems.

\section{Discussion}

\noindent
The few hundred multiple planet candidates discovered by the {\it
  Kepler} spacecraft provide a unique window into
how planets form and evolve, although we do not yet know how to
interpret the limited view that we have through this window. This
paper is focused on the distribution of planets near mean-motion
resonances in multi-planet systems, which has two important features.

First, the number of systems near mean-motion resonances is small: the
left panel of Figure \ref{fig:sim+obs} shows that the only resonance
feature that is significant at more than the 1.5--$\sigma$ level is
the peak at the 3\,:\,2 resonance. As discussed in \S\ref{sect:intro},
this result is striking because since planet
migration is believed to be a common process, and capture into a
first-order resonance during convergent migration is certain if the
eccentricities and inclinations of the planet pair are sufficiently
small and migration is sufficiently slow when the resonance is crossed.

Second, at both the 3\,:\,2 and 2\,:\,1 first-order resonances the
peak in the period-ratio distribution appears just wide of the
resonance (i.e., at period ratios greater than the resonance value, if
the period ratio is defined to be $>1$), rather than at the
resonance. In addition, the 2\,:\,1 resonance appears to have a trough
in the period-ratio distribution just narrow of the resonance,
although the statistical significance of the trough is weak. All of
these features were pointed out already by \cite{FL12}.

Given these findings, it is natural to ask what distribution of period
ratios should be expected near resonance if there is no migration,
i.e., if planets form {\em in situ}. We have investigated two extreme
limits of this process, in which the planet forms fast or slowly
(relative to the characteristic dynamical time associated with the
resonance). We find that in both cases the distribution of test
particles near resonance develops a shape characterized by a peak wide
of the resonance (period  $\mathcal{P}>\mathcal{P}_{\rm res}$
where the period ratio at resonance $\mathcal{P}_{\rm res}>1$) and a
trough narrow of the resonance (see Figures \ref{fig:sudden} and
\ref{fig:slow}). The peak and trough result from a redistribution of
particles from period ratios $\mathcal{P}\lesssim \mathcal{P}_{\rm
  res}$ to $\mathcal{P}\gtrsim \mathcal{P}_{\rm res}$, rather than
from an overall gain or loss of particles as would occur through
processes such as ejection or migration.

\subsection{Planet masses in the {\it Kepler} sample}

\label{sec:typical}

\noindent
The mean and median radius for the planets in the 242 two-planet
systems in \cite{FL12} are $2.46R_\oplus$ and $2.17R_\oplus$,
respectively. To compare our models to the data, i.e., to estimate the
strength parameter $s$ in Figure \ref{fig:p_values}, we need the
mass-radius relation for the {\it Kepler} planets. 

(i) Fitting to the planets in the solar system, \cite{FL12} find
$M=M_\oplus(R/R_\oplus)^{2.06}$ for $R>R_\oplus$, which yields
$M=6.4M_\oplus$ and $4.9M_\oplus$ for the mean and median mass in the
{\it Kepler} two-planet sample. (ii) From a statistical fit to transit
timing variations in the {\it Kepler} sample, \cite{wl12} find
$M=3M_\oplus(R/R_\oplus)$, which implies $7.4M_\oplus$ and
$6.5M_\oplus$ for the mean and median mass. (iii) There are four
planets with measurements of both mass and radius in the radius range
2--$3R_\oplus$; these have mean and median masses of $7.0M_\oplus$ and
$6.1M_\oplus$. All three of these crude approaches suggest a typical
planet mass of about 6--$7M_\oplus$ for the {\it Kepler} planets. 

\subsection{Planet masses required to explain features in the period
  distribution near resonances}

\noindent
Our simple theoretical model, based on slow growth of the planet
masses with no migration, is consistent with the data ($p>0.1$) shown
in Figure \ref{fig:p_values} for values of the strength $s$ between
0.006 and 0.018\footnote{We recognize that the KS test is designed for
  hypothesis testing, not parameter fitting, but the data and the
  models do not justify a more sophisticated approach.}. Equation (\ref{eq:sdef}) gives $s=0.023(m_{\rm pl}/M_J)^{2/3}$
and $s=0.0179(m_{\rm pl}/M_J)^{2/3}$ at 3\,:\,2 and 2\,:\,1, respectively;
focusing on the 3\,:\,2 resonance, which contains the strongest
signal, consistency then requires $40 \lesssim m_{\rm pl}/M_\oplus \lesssim
220$. Our calculations are  for a system containing a test particle and a planet
of mass $m_{\rm pl}$; in the {\it  Kepler} two-planet systems both planets have
non-zero mass and in the absence of a detailed numerical study of this
case the most appropriate choice is probably to identify $m_{\rm pl}$ in
these formulae with twice the mean planet mass. 
Thus the mean planet mass required to explain the resonant structure 
seen in the {\it Kepler} data at the 3\,:\,2 resonance is roughly
20--$100M_\oplus$.

The low end of this mass range is about three times the mean planet
mass estimated in the preceding subsection. There are several
possibilities for bridging this gap:

\paragraph{Test-particle calculations underestimate the dynamical
  effect of resonances} Our analytic calculations are only possible
because one of the two planets was approximated as a test particle. If
both planets have non-zero masses, the dynamical problem acquires
extra degrees of freedom; even the planar single-resonance model has
two degrees of freedom which allows chaos caused by overlap of the
closely spaced resonances with critical angles
$(p+1)\lambda_2-p\lambda_1-\omega_1$ and
$(p+1)\lambda_2-p\lambda_1-\omega_2$.
We have carried out exploratory integrations in which we replace the
three-body system containing a planet of mass $m_{\rm pl}$ and a test
particle with two planets of mass $m_{\rm pl}/2$, leaving the other initial
conditions and parameters the same. We find that the peaks and troughs at the
$3\,:\,2$ and $2\,:\,1$ resonances become significantly stronger, and that a secondary
peak develops narrow of the $2\,:\,1$ resonance. 

\paragraph{Long-term evolution} Systems with more than two degrees of
freedom can exhibit slow evolution due to weak chaos (a classic
example is the Kirkwood gaps in the asteroid belt, which can evolve on
Gyr timescales; \citealt{morb96}). Integrations that follow two
planets with non-zero masses, eccentricities, and inclinations may
have resonant features that evolve over timescales much longer than
those we have examined here. 

\paragraph{Underestimated planet masses}
Simultaneous measurements of mass and radius are available for 8
planets in the radius range 1--$3R_\oplus$. These show an order of
magnitude range of mean density, from $0.7\hbox{$+0.7 \atop
  -0.4$}\mbox{\,g cm}^{-3}$ for Kepler 11f \citep{liss11} to
$8.8\hbox{$+2.2 \atop -2.9$}\mbox{\,g cm}^{-3}$ for Kepler 10b
\citep{bat11}. This wide range implies either a diverse set of
planetary compositions \citep{wl12} or unmodeled errors in some or all
of the measurements. In either case the actual value of the
``typical'' {\it Kepler} planet mass is far more uncertain than the
simple estimate of 6--$7M_\oplus$ obtained above. In particular, if there is a
substantial population of rock-iron planets with radii of
2--$3R_\oplus$ these would have much larger masses: for a radius of
$2.3R_\oplus$, \cite{seager07} find masses of 30--$200M_\oplus$ for a
variety of rock-iron compositions, and \cite{swift12} find masses from
$30M_\oplus$ for basalt to $100M_\oplus$ for nickel-iron planets.

\subsection{Relation to previous work}

\label{sec:previous}

\noindent
\citet{bm12} and \citet{lw12} have shown that tidal dissipation can repel
planet pairs from exact resonance, producing a peak wide of the
resonance and a trough narrow of the resonance as observed in the {\it
  Kepler} data. However, this process requires that (i) the planet
pairs form in resonance; (ii) the correct amount of dissipation is present
to repel the planet pairs from resonance by a percent or so. 

Moreover, if tidal dissipation from the host star were responsible
for shaping the period-ratio distribution around resonances, one should expect
a strong dependence on orbital radius or period (the characteristic
timescale for changes in semi-major axis $a$ varies as $a^8$ in the
tidal evolution model of \citealt{hut81}). However, \citet{rein12b}
points out that the period-ratio distribution in the  {\it Kepler} sample 
restricted to inner-planet periods $<5\,\mbox{d}$ looks identical to
the distribution restricted to inner-planet periods $>5\,\mbox{d}$.
We have repeated Rein's analysis for the restricted sample of two-planet
systems used here and observe that there is a hint that the resonant features
are more pronounced at the 3\,:\,2 resonance for the  
longer period planets, and more pronounced at the 2\,:\,1 resonance
for the shorter periods planets. However, the number of planet
pairs in our restricted sample is too small to make a statistically
significant detection of any difference. 

\citet{rein12b} has also investigated whether stochastic migration can
reproduce the {\it Kepler} period-ratio distribution. He argues that
conventional ``smooth'' migration cannot reproduce the observations
because it produces too many planet systems in exact resonance.
However, these large pile-ups at resonance are smeared out by including
stochastic forces that might be expected from the likely turbulent
nature of the protoplanetary disk.  He shows that for the correct combination
of stochastic and smooth migration forces, the final period-ratio
distribution looks similar to that of the {\it Kepler} planets.

We note that eccentricity measurements of near-resonant planets do not
distinguish between these mechanisms: tidal dissipation damps the free
eccentricity so the residual eccentricity is equal to the forced
eccentricity, which is just the fixed point of the single-resonance
Hamiltonian shown in Figure \ref{fig:fixedpoints}. Slow growth of the
planet causes a test particle on an initially circular orbit to follow
the fixed point as the planet mass grows, so that it will also have
zero free eccentricity unless it has crossed the separatrix.

\section{Summary}

\noindent
We have studied the orbital distribution of two-planet systems near
first-order mean-motion resonances in the simplest possible model of
planet formation: there is no energy dissipation or migration, and
planets form {\em in situ} on circular, coplanar orbits, with masses
growing at a prescribed rate.  We have examined whether this toy model
can explain the signatures in the period-ratio distribution near
resonances that are observed in the {\it Kepler} sample of 242
two-planet systems.

Our approach is to construct a simplified Hamiltonian that isolates
the perturbations due to a given first-order resonance. We then solve
for the long-term dynamics of a test particle in this Hamiltonian,
focusing on two limiting cases: rapid and slow mass growth of the
planet.  We have used numerical integrations of the restricted
three-body problem to confirm that the resonance Hamiltonian
captures the main features of the orbit evolution.  

We find that the distribution in period ratios resembles a ``P-Cygni'' profile,
where orbits are evacuated from narrow of the resonance (i.e., closer
to the perturbing planet) and pile up in regions wide of the
resonance.  These features are present whether the planet grows fast
or slowly, though they are stronger in the latter case, and hence
should be present for a wide range of growth histories.  The resulting
structure in the period-ratio distribution is strongly reminiscent of
the peak-trough feature observed at the $3\,:\,2$ and $2\,:\,1$
resonances in the {\it Kepler} systems.

The model and observed period-ratio distributions near these
resonances are consistent for mean planet masses in the range
20--$100M_\oplus$. This is larger than expected from the handful of {\it
  Kepler} planets with typical radii of 2--$3R_\oplus$ and measured
masses by a factor of 3--15. However, the mass-radius relation for
{\it Kepler} planets is quite uncertain and shows large scatter, and
the required masses are consistent with the masses expected for planets in this
radius range that are composed of a mixture of silicates and
iron with no extended atmosphere. Other effects such as weak chaos operating on Gyr timescales may
also narrow the gap between the model and the observations. Our model
suggests that the resonant features seen in the {\it Kepler} multi-planet
systems may not require either dissipation or migration during the
planet-formation process. 

\acknowledgements 

We are grateful to Subo Dong, Darin Ragozzine, and Hanno Rein for
their insights and suggestions.
RM gratefully acknowledges support from the Institute for 
Advanced Study, and CP acknowledges support from the 
CONICYT Biccentenial  Becas Chile fellowship. 

\appendix

\section{Migration speed and capture into resonance}

\noindent
Capture into a $(p+1)\,:\,p$ resonance during convergent migration is
certain if the planets cross the resonance slowly enough and their
eccentricities and inclinations are small enough. These statements can
be quantified using the formalism of \S\ref{sect:analytic}.

In convergent migration, the dimensionless resonance distance $\Delta$
(eq.\ \ref{eq:delta}) is increasing with time. Since the resonant
Hamiltonian $K$ (eq.\ \ref{e:Kres}) is dimensionless, resonance
crossing without capture should occur when $d\Delta/d\tau > g$ where
$g$ is some constant of order unity. Converting this inequality to
physical units yields
\begin{equation}
\frac{d}{dt}\log n_{\rm pl} > \frac{3^{5/3}g}{4}
|p^2(p+1)|^{1/9}n_{\rm pl}\left|\frac{m_{\rm pl}}{m_*}f_p\right|^{4/3}.
\end{equation}
Define the migration time of the massive planet to be $t_{\rm
  mig}\equiv (d\log n_{\rm pl}/dt)^{-1}$; then the condition for resonance
crossing without capture is
\begin{equation}
t_{\rm mig}  < \frac{3.0\times
10^4\mbox{\,yr}}{g w_p}\frac{P_{\rm pl}}{100\mbox{\,d}}\left(\frac{10M_\oplus}{m_{\rm pl}}\frac{m_*}{M_\odot}\right)^{4/3}
\end{equation}
where $P_{\rm pl}$ is the massive planet's orbital period and $w_p\equiv
|p^2(p+1)|^{1/9}f_p^{4/3}$ is 1.363 for $p=1$ ($2\,:\,1$ resonance)
and 3.377 for $p=2$ ($3\,:\,2$ resonance). The constant $g$ is
estimated to be 2.5 by \cite{fried01} from approximate analytic
calculations and 2.7 by \cite{quil06} from numerical orbit
integrations. 

The maximum eccentricity at which capture into resonance is certain
during convergent migration has been derived by several authors
\citep[e.g.,][]{hl83,bg84}. If the eccentricity of the test particle when far from
resonance is $e_0$, then the area enclosed by its orbit in the $x$-$y$
plane is $A_0=\pi e_0^2/s_e^2$ (eq.~\ref{eq:sedef}).
Because of adiabatic
invariance, this area is conserved as the planet migrates. The test-particle orbit
will be interior to the homoclinic orbit when it first appears, at
$\Delta=1$ (cf.\ Figure \ref{fig:topology}), if $A_0< 6\pi$, the
area of the homoclinic orbit. As the planet continues to migrate and
$\Delta$ continues to grow, the area inside the homoclinic orbit grows
so the test particle remains inside the homoclinic orbit, i.e., it
remains captured in the resonance.  Thus the condition that resonance
capture is certain is $e_0< \sqrt{6}s_e$ or
\begin{equation}
e_0 <
\frac{2^{1/2}3^{1/6}}{|p^2(p+1)|^{2/9}}\left|\frac{m_{\rm pl}}{m_*}f_p\right|^{1/3}
= 0.0528 h_p\left(\frac{m_{\rm pl}}{10M_\oplus}\frac{M_\odot}{m_*}\right)^{1/3},
\end{equation}
where $h_p=|p^2(1+p)|^{-2/9}|f_p|^{1/3}$ is 
0.909 for $p=1$ and 0.728 for $p=2$.


\begin{thebibliography}{}

\bibitem[Anglada-Escud{\'e} et al.(2010)]{ae10} 
Anglada-Escud{\'e}, G., L{\'o}pez-Morales, M., 
\& Chambers, J.~E.\ 2010, \apj, 709, 168 

\bibitem[Baluev(2011)]{bal11} Baluev, R.~V.\ 2011, Celestial 
Mechanics and Dynamical Astronomy, 111, 235 

\bibitem[Batalha et al.(2011)]{bat11} Batalha, N.~M., 
Borucki, W.~J., Bryson, S.~T., et al.\ 2011, \apj, 729, 27 

\bibitem[Batygin \& Morbidelli(2012)]{bm12} Batygin, K., \&
  Morbidelli, A.\ 2012, arXiv:1204.2791 

\bibitem[Borderies \& Goldreich(1984)]{bg84} Borderies, N., \&
  Goldreich, P.\ 1984, Celestial Mechanics, 32, 127


\bibitem[Delisle et al.(2012)]{las12} Delisle, J.-B., Laskar, 
J., Correia, A.~C.~M., \& Bou{\'e}, G.\ 2012, \aap, 546, A71 


\bibitem[Fabrycky et al.(2012)]{FL12} Fabrycky, D.~C., 
Lissauer, J.~J., Ragozzine, D., et al.\ 2012, arXiv:1202.6328 


\bibitem[Friedland(2001)]{fried01} Friedland, L.\ 2001, \apjl, 
547, L75 

\bibitem[Henrard 
\& Lema\^itre(1983)]{hl83} Henrard, J., \& Lema\^itre, A.\ 1986,
Celestial Mechanics, 30, 197


\bibitem[Hut(1981)]{hut81} Hut, P.\ 1981, \aap, 99, 126 

\bibitem[Lecar et al.(2001)]{Lecar:2001}
Lecar, M., Franklin, F. A., Holman, M. J., \& Murray, N. J. 2001, \araa, 39,
581

\bibitem[\protect\citeauthoryear{Liou \& Malhotra}{Liou \&
    Malhotra}{1997}]{Liou:1997} Liou, J.C., \& Malhotra, R. 1997,
  Science, 275, 375 

\bibitem[Lissauer et al.(2011a)]{liss11} Lissauer, J.~J., 
Fabrycky, D.~C., Ford, E.~B., et al.\ 2011a, \nat, 470, 53 

\bibitem[Lissauer et al.(2011b)]{LF11} Lissauer, J.~J., 
   Ragozzine, D., Fabrycky, D.~C., et al.\ 2011b, \apjs, 197, 8 

\bibitem[Lissauer et al.(2012)]{liss12} Lissauer, J.~J., 
Marcy, G.~W., Rowe, J.~F., et al.\ 2012, \apj, 750, 112 

\bibitem[Lithwick \& Wu(2012)]{lw12} Lithwick, Y., \& Wu, Y.\ 2012, 
\apj, 756, L11

\bibitem[Malhotra(1993)]{mal93} Malhotra, R.\ 1993, \nat, 
365, 819 

\bibitem[Malhotra(1994)]{mal94} Malhotra, R.\ 1994, Physica D, 77, 289 

\bibitem[Malhotra(1998)]{mal98} Malhotra, R.\ 1998, in Solar 
System Formation and Evolution, eds. D.\ Lazzaro et al., PASP Conference Series 149, 37 

\bibitem[Malhotra et al.(1992)]{mal92} Malhotra, R., Black, 
D., Eck, A., \& Jackson, A.\ 1992, \nat, 356, 583 

\bibitem[\protect\citeauthoryear{Minton \& Malhotra}{Minton \&
 Malhotra}{2009}]{Minton:2009}
Minton, D.~A.,  \& Malhotra, R. 2009, Nature, 457, 1109

\bibitem[\protect\citeauthoryear{Minton \& Malhotra}{Minton \&
 Malhotra}{2010}]{Minton:2010}
Minton, D.~A.,  \& Malhotra, R. 2010, Icarus, 207, 744

\bibitem[Morbidelli(1996)]{morb96} Morbidelli, A.\ 1996, \aj, 
111, 2453 

\bibitem[Murray \& Dermott(1999)]{md99} Murray, C.~D., \& Dermott,
  S.~F.~1999, Solar System Dynamics (Cambridge: Cambridge University
  Press). 

\bibitem[\protect\citeauthoryear{Murray \& Holman}{Murray \&
 Holman}{1997}]{Murray:1997}
Murray, N., \& Holman, M. 1997, \aj, 114, 1246

\bibitem[Quillen(2006)]{quil06} Quillen, A.~C.\ 2006, \mnras, 
365, 1367 

\bibitem[Rasio et al.(1992)]{ras92} Rasio, F.~A., Nicholson, 
P.~D., Shapiro, S.~L., \& Teukolsky, S.~A.\ 1992, \nat, 355, 325 

\bibitem[Rein(2012)]{rein12b} Rein, H. 2012, \mnras, 427, L21

\bibitem[Rein et al.(2012)]{rein12} Rein, H., Payne, M.~J., 
Veras, D., \& Ford, E.~B.\ 2012, \mnras, 426, 187

\bibitem[Seager et al.(2007)]{seager07}
Seager, S., Kuchner, M., Hier-Majumder, C.~A.,
\& Militzer, B., 2007, \apj, 669, 1279

\bibitem[Snellgrove et 
al.(2001)]{spn01} Snellgrove, M.~D., Papaloizou, J.~C.~B., \& Nelson, R.~P.\ 2001, \aap, 374, 1092 

\bibitem[Swift et al.(2012)]{swift12} Swift, D.~C., Eggert, 
J.~H., Hicks, D.~G., et al.\ 2012, \apj, 744, 59 

\bibitem[Tanaka et al.(2002)]{ttw02} Tanaka, H., Takeuchi, 
T., \& Ward, W.~R.\ 2002, \apj, 565, 1257 

\bibitem[Tremaine \& Dong(2011)]{TD11} Tremaine, S., \& Dong, S.\ 2011,
\aj, 143, 94

\bibitem[Wisdom(1980)]{wisdom:1980} Wisdom, J., 1980, AJ, 85, 1122

\bibitem[Wisdom(1983)]{wis83} Wisdom, J.\ 1983, Icarus, 56, 
51 


\bibitem[Wright et al.(2011)]{wright} Wright J.~T. et al. 2011, PASP,
  123, 412

\bibitem[Wu \& Lithwick(2012)]{wl12} Wu, Y., \& Lithwick, Y.\ 2012, arXiv:1210.7810

\end{thebibliography}
\end{document}